# Refinement and Verification of Real-Time Systems[†]

Paul Z. Kolano[1], Carlo A. Furia[2], Richard A. Kemmerer[3], and Dino Mandrioli[4]

[1] NASA Advanced Supercomputing Division,
NASA Ames Research Center, Moffett Field, CA 94035 U.S.A.
**kolano@nas.nasa.gov**

[2] Chair of Software Engineering, Department of Computer Science
ETH Zurich, ETH Zentrum, 8092 Zürich, Switzerland
**caf@inf.ethz.ch**

[3] Reliable Software Group, Computer Science Department,
University of California, Santa Barbara, CA 93106 U.S.A.
**kemm@cs.ucsb.edu**

[4] Dipartimento di Elettronica e Informazione,
Politecnico di Milano, Milano 20133, Italy
**mandriol@elet.polimi.it**

**Abstract**

This paper discusses highly general mechanisms for specifying the refinement of a real-time system as a collection of lower level parallel components that preserve the timing and functional requirements of the upper level specification. These mechanisms are discussed in the context of ASTRAL, which is a formal specification language for real-time systems. Refinement is accomplished by mapping all of the elements of an upper level specification into lower level elements that may be split among several parallel components. In addition, actions that can occur in the upper level are mapped to actions of components operating at the lower level. This allows several types of implementation strategies to be specified in a natural way, while the price for generality (in terms of complexity) is paid only when necessary. The refinement mechanisms are first illustrated using a simple digital circuit; then, through a highly complex phone system; finally, design guidelines gleaned from these specifications are presented.

## 1. Introduction

Refinement is a fundamental design technique that has often challenged the "formal methods" community. In most cases, mathematical elegance and proof manageability have exhibited a deep trade-off with the flexibility and freedom that are often needed in practice to deal with unexpected or critical situations. A typical example is provided by algebraic approaches that exploit some notion of homomorphism between algebraic structures. When applied to parallel systems, such approaches led to the notion of observational equivalence of processes [HM 85] (i.e. the ability of the lower level process to exhibit all and only the observable behaviors of the upper level one). Observational equivalence, however, has been proved too restrictive to deal with general cases and more flexible notions of inter-level relations have been advocated [DiM 99] (see also Section 11).

---

[†] A preliminary version of this paper appeared as "Parallel Refinement Mechanisms for Real-Time Systems" in *Proc. 3rd Int. Conf. on Fundamental Approaches to Software Engineering*. Springer, Berlin, 2000, pp. 35-50.

The issue of refinement becomes even more critical when dealing with real-time systems where time analysis is a crucial factor. In fact, not surprisingly, whereas the literature exhibits a fairly rich set of contributions to the refinement of traditional, non-real-time systems, to the best of our knowledge, only a few, relatively limited proposals have been published for the formal refinement of real-time systems. We discuss them in Section 11.

In this paper, we propose highly general refinement mechanisms that allow several types of implementation strategies to be specified in a natural way. In particular, processes can be implemented both sequentially, by refining a single complex transition as a sequence or selection of more elementary transitions, and in a parallel way, by mapping one process into several concurrent ones. This allows one to increase the amount of parallelism through refinement whenever needed or wished.

In our refinement framework, even *asynchronous implementation policies* are allowed, in which lower level actions can have durations unrelated to upper level ones, provided that their effects are made visible in the lower level at times that are compatible with those specified by the upper level. In other words, the refinement is sound if it preserves the (timing) requirements in the upper level. In practice, this is achieved by specifying the durations of transitions in the lower level in such a way that the time constraints of the requirements are not violated. For instance, in a phone system, many calls must be served simultaneously, possibly by exploiting concurrent service by many processors. Such services, however, are asynchronous since calls occur in an unpredictable fashion at any instant. Therefore, it is not easy to describe a call service that manages a set of calls within a given time interval in an abstract way that can be naturally refined as a collection of many independent and individual services of single calls, possibly even allowing a dynamic allocation of servers to the phones issuing the calls. In Section 9, we describe how this goal can be achieved by applying the mechanisms described in this paper. In Section 11, we analyze several works dealing with the problem of refinement for real-time systems, and we show in which aspects they fall short of the approach in the present paper.

Not surprisingly, generality has a price in terms of complexity. In our approach, however, this price is paid only when necessary. Simple implementation policies yield simple specifications, whereas complex specifications are needed only for sophisticated implementation policies. The same holds for the proof system, which is built hand-in-hand with the implementation mechanisms.

This work is presented in the context of ASTRAL, which is a formal specification language for real-time systems with the following distinguishing features:

- It is rooted in both ASLAN [AK 85], an untimed state machine formalism, and TRIO [GMM 90], a real-time temporal logic, yielding a new, logic-based, process-oriented specification language. A real-time system is modeled by one or more instances of multiple process types.

- It has composable modularization mechanisms that allow a complex system to be built as a collection of interacting processes. It also has refinement mechanisms to construct a process as a sequence of layers, where each layer is the implementation of the layer above. This does not confine the user to a strictly top-down development process but is liberal enough to permit a mixed top-down and bottom-up approach to development with the unavoidable trials and errors. However, the *presentation* of the mechanisms and the documentation resulting from its application are strictly top-down to enhance clarity and to summarize the overall achievements, as it is clearly advocated in [PC86] and now fairly customary.

- It has a proof obligation system that allows one to formally prove properties of interest as consequences of process specifications. This proof system is incremental since complex proofs of complex systems can be built by composing small proofs that can be carried out, for the most part, independently of each other. ASTRAL's proofs are of two types. Intra-level proofs guarantee system properties on the basis of local properties that only refer to a single process type. Inter-level proofs guarantee that layer i+1 is a correct implementation of layer i without the need to redo intra-level proofs from scratch.

- It is supported by a tool suite that supplies the typical functions of a formal method such as ASTRAL: editing and managing specifications, deriving proof obligations and supporting their proof in a semiautomatic way (of course, given its generality, ASTRAL is undecidable).

In this paper, we resume the issue of ASTRAL layering mechanisms and the inter-level proofs, which were addressed in a preliminary and fairly restrictive way in [CKM 95].

This paper is structured as follows. Section 2 provides the necessary background on the ASTRAL language. Sections 3 and 4 summarize previous purely sequential refinement mechanisms and the proof obligations for the sequential mechanisms, respectively. Section 5 motivates the need for their extensions through a simple running example and illustrates the generalized and parallel refinement mechanisms. Section 6 shows their application to the running example. Section 7 presents the proof obligations needed to guarantee implementation correctness and Section 8 applies them to the running example. In Section 9, a much more complex application of the parallel refinement mechanisms is demonstrated by refining the central controller of a telephone system. Section 10 proposes some methodological guidelines supporting the refinement process of complex systems, which we derived from our experience on several case studies. Section 11 provides an extensive summary of several related works dealing with the issue of refinement, comparing them with our own framework. It also provides an assessment of the generality and usability of the techniques and methodology of this paper based on the grounds of the experience with the phone system case study. Finally, Section 12 provides some concluding remarks.

This paper focuses attention on the essential conceptual aspects of linguistic, proof, and methodological features of ASTRAL. Its contents should be self-contained and fully understandable with no further reading. All remaining technical details (complete specifications and proofs, further examples, etc.) are fully reported in [Kol 99], which also describes ASTRAL's tool suite.

## 2. ASTRAL Overview

An ASTRAL system specification is comprised of a single global specification and a collection of state machine specifications. Each state machine specification represents a process type of which there may be multiple, statically generated, instances. The *global specification* contains declarations for the process types that comprise the system, types and constants that are shared among more than one process type, and assumptions about the global environment and critical requirements for the whole system.

An ASTRAL *process specification* consists of a sequence of *levels*. Each level is an abstract data type view of the process being specified. The first ("top level") view is a very abstract model of what constitutes the process (types, constants, variables), what the process does (state transitions), and the critical requirements the process must meet (invariants and schedules). Lower levels are increasingly more detailed with the lowest level corresponding closely to high level code.

The process being specified is thought of as being in various *states*, where one state is differentiated from another by the values of the *state variables*, which can be changed only by means of *state transitions*. Transitions are specified in terms of entry and exit assertions, where *entry assertions* describe the constraints that state variables must satisfy in order for the transition to fire, and *exit assertions* describe the constraints that are fulfilled by state variables after the transition has fired. An explicit non-null duration is associated with each transition. Transitions are executed as soon as they are enabled if no other transition is executing in that process.

Every process can export both state variables and transitions. Exported variables are readable by other processes while exported transitions are callable from the external environment. Inter-process communication is accomplished by inquiring about the values of exported variables and the start and end times of exported transitions. The *export* clause of a process specification indicates which variables and transitions that process exports. The *import* clause indicates which variables and transitions exported by other processes and which types and constants declared in the global specification are referenced by the process.

In addition to specifying system state (through process variables and constants) and system evolution (through transitions), an ASTRAL specification also defines system critical requirements and assumptions about the behavior of the environment that interacts with the system. Assumptions about the behavior of the environment are expressed by means of *environment clauses* that describe the pattern of calls to external transitions, which are the stimuli to which the system reacts. Critical requirements are expressed by means of *invariants* and *schedules*. Invariants represent requirements that must hold in every state reachable from the initial state, no matter what the behavior of the external environment is, while schedules represent additional properties that must be satisfied provided that the external environment behaves as assumed. Optional *axiom clauses* can be specified to assert properties of types and constants, which can be used in the proofs of the critical requirements.

Invariants and schedules are proved over all possible executions of a system. A system execution is a set of process executions that contains one process execution for each process instance in the system. A process execution for a given process instance is a history of events on that instance. The value of an expression E at a time t1 in the history can be obtained using the *past* operator, past(E, t1), such that t1 ≤ now, where now is a special global variable used to denote the current time in the system. There are four types of events that may occur in an ASTRAL history. A *call event* occurs for an exported transition tr1 at a time t1 if and only if tr1 was called from the external environment at t1. A *start event* occurs for a transition tr1 at a time t1 if and only if tr1 fires at t1. Similarly, an *end event* occurs if tr1 ends at t1. Finally, a *change event* occurs for a variable v1 at a time t1 if and only if v1 changes value at t1. Note that change events can only occur when an end event occurs for some transition. The corresponding predicates Call(tr1, t1), Start(tr1, t1), End(tr1, t1), and Change(v1, t1) are true at the time now if and only if t1 was the last time at which the given event occurred (i.e. no event of the same type has occurred at any time t such that t1 < t ≤ now). An introduction and complete overview of the ASTRAL language can be found in [CGK 97].

The example system used throughout the remainder of the paper is shown in Figure 2. This system is a circuit that computes the value of a * b + c * d, given inputs a, b, c, and d. The ASTRAL specification for the circuit is shown below.

```
GLOBAL SPECIFICATION MA_Circuit
   PROCESSES
      the_mult_add: Mult_Add
   TYPE
      pos_real: TYPEDEF r: real (r > 0)

PROCESS Mult_Add
   EXPORT
      compute, output
   IMPORT
      pos_real
   CONSTANT
      dur1: pos_real
   VARIABLE
      output: integer
   AXIOM
      TRUE
   INVARIANT
      FORALL t1: time, a, b, c, d: integer
         ( Start(compute(a, b, c, d), t1)
      →    FORALL t2: time
            ( t1 + dur1 ≤ t2 & t2 ≤ now
      →       past(output, t2) = a * b + c * d))
   TRANSITION   compute(a,b,c,d: integer)
      ENTRY  [TIME:   dur1]
         TRUE
      EXIT
         output = a * b + c * d
```

The MA_Circuit system consists of one process type specification: Mult_Add. The *processes* declaration, which occurs in the global specification, declares that there is one instance of Mult_Add. The Mult_Add process imports the global type pos_real, which is defined to be the real numbers greater than zero using the *typedef* operator. The Mult_Add process has a single transition, compute, which when called from the external environment with four parameters, sets the integer variable output to the sum of the products of the first and last pair of parameters after duration dur1 time has elapsed. The invariant clause of Mult_Add states that if the last start of compute occurred at least dur1 time in the past, then output will contain the appropriate sum of products from dur1 after the last start of compute up until the current time. The axiom clause of Mult_Add is empty (i.e. TRUE).

## 3. Sequential Refinement Mechanisms

Refinement mechanisms for ASTRAL were defined in [CKM 95]. In this definition, an ASTRAL process specification consists of a sequence of levels where the behavior of each level is implemented by the next lower level in the sequence. Given two ASTRAL process level specifications $P_U$ and $P_L$, where $P_L$ is a refinement of $P_U$, the implementation statement, hereafter referred to as the IMPL mapping, defines a mapping from all the types, constants, variables, and transitions of $P_U$ into their corresponding terms in $P_L$, which are referred to as *mapped* types, constants, variables, or transitions. The IMPL mapping is extended in the "natural way" to expressions (e.g. IMPL(a * b) ≡ IMPL(a) * IMPL(b)). $P_L$ can also introduce types, constants and/or variables that are not mapped, which are referred to as the *new* types, constants, or variables of $P_L$. Note that $P_L$ cannot introduce any new transitions (i.e. each transition of $P_L$ must be a mapped transition). A transition of $P_U$ can be mapped into a sequence of transitions, a selection of transitions, or any combination thereof.

A selection mapping of the form $T_U == A_1 \& T_{L.1} | A_2 \& T_{L.2} | ... | A_n \& T_{L.n}$, is defined such that when the upper level transition $T_U$ fires, one and only one lower level transition $T_{L.j}$ fires, where $T_{L.j}$ can only fire when both its entry assertion and its associated "guard" $A_j$ are true. The left side of Figure 1 depicts a selection of transitions.

A sequence mapping of the form $T_U ==$ WHEN $Entry_L$ DO $T_{L.1}$ BEFORE $T_{L.2}$ BEFORE ... BEFORE $T_{L.n}$ OD, defines a mapping such that the sequence of transitions $T_{L.1}; ...; T_{L.n}$ is enabled (i.e. can start) whenever $Entry_L$ evaluates to true. Once the sequence has started, it cannot be interrupted until all of its transitions have been executed in order. The starting time of the upper level transition $T_U$ corresponds to the starting time of the sequence (which is not necessarily equal to the starting time of $T_{L.1}$ because of a possible delay between the time when the sequence starts and the time when $T_{L.1}$ becomes enabled), while the ending time of $T_U$ corresponds to the ending time of the last transition in the sequence, $T_{L.n}$. Note that the only transition that can modify the value of a mapped variable is the last transition in the sequence. This further constraint is a consequence of the ASTRAL communication model. That is, in the upper level, the new values of the variables affected by $T_U$ are broadcast when $T_U$ terminates. Thus, mapped variables of $P_L$ can be modified only when the sequence implementing $T_U$ ends. The right side of Figure 1 depicts a sequence of transitions.

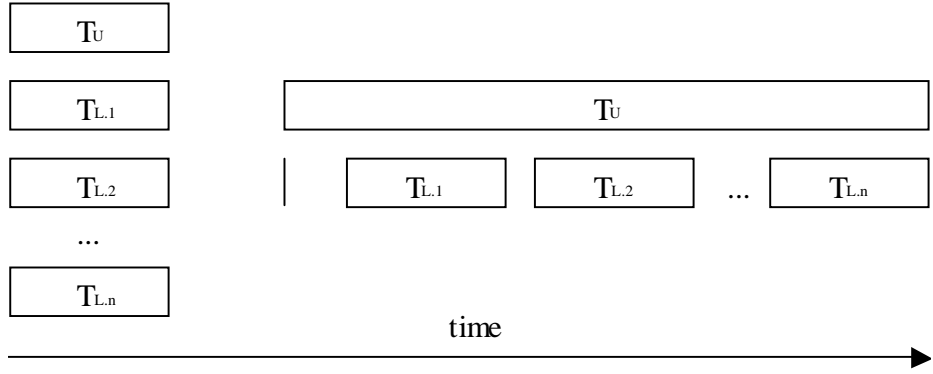

Figure 1: Selection and sequence mappings

## 4. Proof Obligations for Sequential Refinement Mechanisms

The inter-level proofs consist of showing that each upper level transition is correctly implemented by the corresponding sequence, selection, or combination thereof in the next lower level. For selections, it must be shown that whenever the upper level transition $T_U$ fires, one of the lower level transitions $T_{L.j}$ fires, that the effect of each $T_{L.j}$ implies the effect of $T_U$, and that the duration of each $T_{L.j}$ is equal to the duration of $T_U$, where $1 \leq j \leq n$. These obligations are, respectively

- ($S_0$)     IMPL(Entry$_U$) $\leftrightarrow$ $A_1$ & Entry$_{L.1}$ | ... | $A_n$ & Entry$_{L.n}$
- ($S_{1.j}$)    $A_j{'}$ & Entry$_{L.j}{'}$ & Exit$_{L.j}$ $\rightarrow$ IMPL(Exit$_U$)
- ($S_2$)     Dur$_{L.1}$ = Dur$_{L.2}$ = ... = Dur$_{L.n}$ = Dur$_U$

Note that in Astral a prime (') is used to denote the value a variable had when a transition fired. For sequences, it must be shown that the sequence is enabled if and only if $T_U$ is enabled, that the effect of the sequence is equivalent to the effect of $T_U$, and that the duration of the sequence (including any initial delay after Entry$_L$ is true) is equal to the duration of $T_U$. These are shown by the following n+2 incremental proof obligations, where $1 \leq j \leq n$.

- ($P_0$)     IMPL(Entry$_U$) $\leftrightarrow$ Entry$_L$
- ($P_{j+1}$)    past(Entry$_L$, Now - Dur$_U$) & Start($T_{L.1}$, $t_1$) & ... & Start($T_{L.j}$, $t_j$)
         $\rightarrow$ EXISTS $t_{j+1}$: Time
            (   $t_{j+1} \geq t_j$ + Dur$_{L.j}$ & $t_{j+1}$ + $\sum_{k=j+1}^{n}$Dur$_{L.k}$ $\leq$ Now
            & past(Entry$_{L.j+1}$, $t_{j+1}$))
               where in obligation $P_n$, "$\leq$ Now" is replaced by "= Now"
- ($P_{n+1}$)    past(Entry$_L$, Now - Dur$_U$) & past(Exit$_{L.1}$, $t_1$ + Dur$_{L.1}$) & ... & Exit$_{L.n}$
         $\rightarrow$ IMPL(Exit$_U$)

The idea of the selection and sequence obligations is that whenever an upper level transition is enabled, some lower level sequence or selection will be enabled because the entry assertions are equivalent. Similarly, whenever an upper level transition ends, some lower level sequence or selection will end because the durations are the same. Finally, the effect produced by an upper level transition implies the effect produced by the lower level transition, because the IMPL of the exit assertion of the upper level transition holds at the end of the lower level sequence or selection. This means that the upper and lower levels will have equivalent executions.

## 5. Parallel Refinement Mechanisms

In the sequential mechanisms, refinement occurs at the transition level, where the behavior of each upper level transition can be specified in greater detail at the lower level. We now extend the ASTRAL refinement mechanisms to include process level refinement, which allows a process to be refined as a collection of components that operate in parallel. For example, a reasonable refinement of the Mult_Add circuit in Figure 2 is shown in Figure 3. Here, the refinement of the system consists of two multipliers that compute a * b and c * d in parallel and an adder that adds the products together and produces the sum. This refinement cannot be expressed in the sequential mechanisms due to the parallelism between the two multipliers. The new parallel mechanisms introduced below, however, easily expresses this refinement.

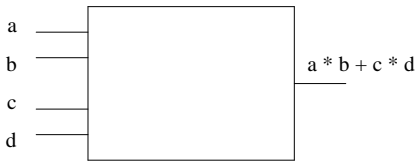 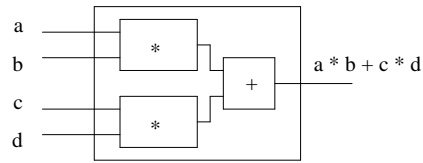

Figure 2: Mult_Add circuit          Figure 3: Refined Mult_Add circuit

In parallel refinement, an upper level transition may be implemented by a dynamic set of lower level transitions. To guarantee that an upper level transition is correctly implemented by the lower level, it is necessary to define the events that occur in the lower level when the transition is executed in the upper level. It must then be shown that these events will only occur when the upper level transition ends and that the effect will be equivalent. Like the sequential refinement mechanisms of [CKM 95], an IMPL mapping is used, which describes how items in an upper level are implemented by items in the next lower level. The items of the upper level include variables, constants, types, and transitions. In addition, the implementation mapping must describe how upper level expressions are transformed into lower level expressions.

### 5.1. Parallel Sequences and Selections

A natural but limited approach to defining parallel transition mappings is to extend the sequential sequence and selection mappings into parallel sequence and selection mappings. Thus, a "||" operator could be allowed in transition mappings, such that "$P_1$.tr1 || $P_2$.tr2" indicates that tr1 and tr2 occur in parallel on processes $P_1$ and $P_2$, respectively. With this addition, the compute transition of the Mult_Add circuit could be expressed as the following.

```
IMPL(compute(a, b, c, d)) == WHEN TRUE DO
    (M1.multiply(a, b) || M2.multiply(c, d))
        BEFORE A1.add(M1.product, M2.product) OD,
```
where M1 and M2 are the multipliers and A1 is the adder.

Although parallel sequences and selections work well for the example, they do not allow enough flexibility to express many reasonable refinements. For example, consider a production cell that executes a produce transition every time unit to indicate the production of an item. In a refinement of this system, the designer may wish to implement produce by defining two "staggered" production cells that each produce an item every two time units, thus effectively producing an item every time unit. The upper level production cell $P_U$ and the lower level production cells $P_{L.1}$ and $P_{L.2}$ are shown in Figure 4. Note that the first transition executed on $P_U$ is init, which represents the "warm-up" time of the production cell in which no output is produced.

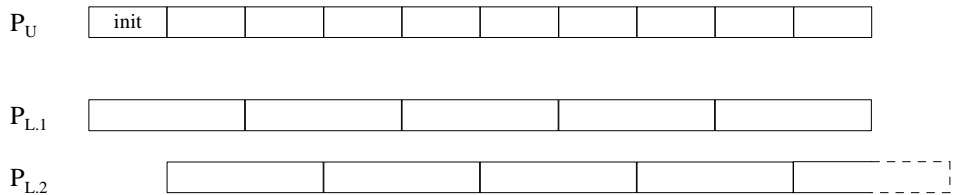

Figure 4: Production cell refinement

This refinement cannot be expressed using parallel sequences and selections, because there is no sequence of parallel transitions in the lower level that corresponds directly to produce in the upper level. When produce starts in the upper level, one of the lower level produce's will start and when produce ends in the upper level, one of the lower level produce's will end and achieve the effect of upper level produce, but the produce that starts is not necessarily the produce that achieves the effect of the corresponding end.

## 5.2. Parallel Start, End, and Call Mappings

The desired degree of flexibility is obtained by using transition mappings that are based on the start, end, and call of each transition. For each upper level transition $T_U$, a start mapping "IMPL(Start($T_U$, now)) == $Start_L$" and an end mapping "IMPL(End($T_U$, now)) == $End_L$" must be defined. If $T_U$ is exported, a call mapping "IMPL(Call($T_U$, now)) == $Call_L$" must also be defined. These mappings are defined with respect to the current time, now, hence are defined for all possible times.

Here, $Start_L$, $End_L$, and $Call_L$ are well-formed formulas using lower level transitions and variables. For the most part, the end and call mappings will correspond to the end or call of some transition in the lower level, whereas the start mapping may correspond to the start of some transition or some combination of changes to variables, the current time, etc. Call mappings are restricted such that for every lower level exported transition $T_L$, Call($T_L$) must be referenced in some upper level exported transition call mapping IMPL(Call($T_U$, now)). This restriction expresses the fact that the interface of the process to the external environment cannot be changed. For parameterized transitions, only the call mapping may reference the parameters given to the transition. Any parameter referenced in a call mapping must be mapped to a call parameter of some lower level transition and the corresponding start mapping must contain the same transitions as the call mapping. Thus, the start and end parameters are taken from the associated set of unserviced call parameters.

With these mappings, the initialize and produce transitions can be mapped as follows. Note that the parallelism between $P_{L.1}$ and $P_{L.2}$ is implied by the overlapping start and end times of produce on each process and not by a built-in parallel operator.

```
IMPL(Start(initialize, now)) ==            IMPL(End(initialize, now)) ==
    now = 0 & P_{L.1}.Start(produce, now)      now = 1
IMPL(Start(produce, now)) ==               IMPL(End(produce, now)) ==
    IF now mod 2 = 0                           IF now mod 2 = 0
    THEN P_{L.1}.Start(produce, now)           THEN P_{L.1}.End(produce, now)
    ELSE P_{L.2}.Start(produce, now)           ELSE P_{L.2}.End(produce, now)
    FI                                         FI
```

It is possible to show that any sequence or selection mapping as defined in [CKM 95] can be described by the start and end mappings. For a selection $T_U == A_1\ \&\ T_{L.1}\ |\ A_2\ \&\ T_{L.2}\ |\ ...\ |\ A_n\ \&\ T_{L.n}$, the start of the upper level transition $T_U$ occurs whenever one of the transitions $T_{L.i}$ starts and its associated guard holds. This is described by the start mapping

```
IMPL(Start(T_U, now)) ==
    (   A_1 & P_L.Start(T_{L.1}, now)
    |   A_2 & P_L.Start(T_{L.2}, now)
    |   ...
    |   A_n & P_L.Start(T_{L.n}, now))
```

The end of $T_U$ occurs whenever one of the transitions $T_{L.i}$ ends. This is described by the end mapping

```
IMPL(End(T_U, now)) ==
    (   P_L.End(T_{L.1}, now)
    |   P_L.End(T_{L.2}, now)
    |   ...
    |   P_L.End(T_{L.n}, now))
```

For a sequence $T_U ==$ WHEN $Entry_L$ DO $T_{L.1}$ BEFORE $T_{L.2}$ BEFORE ... BEFORE $T_{L.n}$ OD, the start of $T_U$ occurs whenever the entry condition of the sequence $Entry_L$ holds. This is described by the start mapping

```
IMPL(Start(T_U, now)) == Entry_L
```

The end of $T_U$ occurs whenever the last transition $T_{L.n}$ ends. This is described by the end mapping

```
IMPL(End(T_U, now)) == P_L.End(T_{L.n}, now)
```

Note that nothing is stated about the transitions that occur between when $Entry_L$ holds and $T_{L.n}$ ends. The rest of the sequence will appear in the proof obligations, where it will need to be proved that each transition of the sequence occurs at the proper time to guarantee that the sequence takes $Dur_U$ time and that the effect of $T_U$ is satisfied.

### 5.3. Other Mappings

Besides transitions, the user must also define mappings for types, constants, and variables. For the most part, the constant and variable mappings are similar to the mappings used in [CKM 95]. In [CKM 95], however, the lower level only consisted of a single process type, so variable mappings only referred to the variables of a single process. In the parallel mechanisms, variable mappings can refer to variables and transitions of any process in the refinement. For example, the mapping for a variable x can be defined IMPL(x) == (P1.y + P2.z) / 2. That is, x is the average of the y variable of process P1 and the z variable of process P2. An additional consequence of having multiple lower level processes is that information must be provided about the processes that make up the lower level. Thus, the implementation section contains a processes clause similar to the clause in the global specification that describes all the process instances of the new lower level. The upper level process is the parallel composition of these process instances.

The IMPL mappings for types, constants, and variables are not discussed in [CKM 95], but are assumed to be an extension of the mappings in [AK 85], modified to include ASTRAL constructs. The mappings in [AK 85], however, do not consider any nontrivial type mappings; therefore, they do not allow the IMPL translation of an arbitrary expression to be constructed. For example, consider an upper level with a type "S: set of T" and a variable "v_s: S". In the lower level, the specifier may wish to implement S and v_s as "L: list of T" and "v_l: L", such that if an element of type T is in the set v_s, the element is somewhere on the list v_l. The IMPL mapping can be defined as IMPL(S) == L and IMPL(v_s) == v_l. Suppose an entry assertion $Entry_U$ in a transition of the upper level states that "t ISIN v_s", where t is an element of type T. The proof obligations require IMPL($Entry_U$) be constructed in order to attempt the proofs. There is no mention in [CKM 95] or [AK 85], however, of how to construct the lower level expression for such a type mapping. If only variables are transformed, the entry assertion becomes "IMPL(t) ISIN v_l", but v_l is a list and ISIN is an operator on sets. Thus, it is necessary to define IMPL mappings in much more detail to be able to attempt the proof obligations.

In the example above a set type was mapped to a list type. This created a problem because the set operators are not valid on lists. In general, there is a problem any time an upper level type T is mapped to a lower level type IMPL(T) that is not "compatible" with T. To be more precise, define types T and IMPL(T) to be compatible if and only if:

- T is an undefined type
- T is identical to IMPL(T)
- T is a list of E and IMPL(T) is a list of IE and E is compatible with IE
- T is a set of E and IMPL(T) is a set of IE and E is compatible with IE
- T is a structure of $ID_1$: $E_1$, ..., $ID_n$: $E_n$ and IMPL(T) is a structure of $ID_1$: $IE_1$, ..., $ID_n$: $IE_n$ and $E_i$ is compatible with $IE_i$
- IMPL(T) is a typedef of E and T is compatible with E

Note that type mappings are restricted such that built-in types cannot be mapped and that any alias or subtype of a given supertype can only be mapped if no other alias or subtype has been mapped. For example, the types "T1" and "T2: typedef t1: T1 (P(t1))" cannot both be mapped. In this restriction, the built-in types integer and time are assumed to be subtypes of the supertype real.

Examples of compatible types are:

(1) T: (e1, e2), IMPL(T): (e1, e2)
(2) T: list of real, IMPL(T): list of integer

Examples of incompatible types are:

(1) T: (open, closed), IMPL(T): (open, closed, opening, closing)
(2) T: list of integer, IMPL(T): set of integer
(3) T: list of bool, IMPL(T): integer
(4) T: structure of (i1: integer, i2: integer), IMPL(T): structure of (j1: integer, j2: integer)

Note that type (4) is incompatible due to the use of different identifiers within the structure definitions.

The IMPL mapping describes how items in an upper level are implemented by items in a lower level. The items of the upper level include variables, constants, types, and transitions. In addition, the IMPL mapping must describe how upper level expressions are transformed into lower level expressions. In many cases, namely when variables and constants are mapped to expressions of compatible types, the basic mappings are sufficient to transform upper level expressions into lower level expressions. When mappings occur between incompatible types, however, the basic mappings must be supplemented with additional mapping information.

For each upper level type T that is mapped to an incompatible lower level type IMPL(T) and for each variable or constant of type T that is mapped to a lower level expression of an incompatible type $T_L$, a mapping must be defined for each operator op in the upper level that is used on an item of type T. For simplicity, assume that all operators are in prefix notation.

$$IMPL(op(v_1: T_1, ..., v_i: T_i, ..., v_n: T_n)) == f(IMPL(v_1), ..., IMPL(v_i), ..., IMPL(v_n))$$

The operator mappings are restricted such that none of the timed operators (i.e. start, end, call, change, and past) can be mapped. The start, end, and call operators will always be mapped as a simple replacement mapping as described earlier. The past and change operators will always use the "natural" operator mapping which is defined as follows.

$$IMPL\_0(op(v_1: T_1, ..., v_n: T_n)) == op(IMPL(v_1), ..., IMPL(v_n))$$

In other words, the natural mapping for operators passes the IMPL construct through to its operands. For example, IMPL(past(A, t)) == past(IMPL(A), IMPL(t)) and IMPL(change(A, t)) = change(IMPL(A), IMPL(t)). The implementation of any operator that does not have an explicit mapping for its operand types is defined to be the natural operator mapping.

As an example of an operator mapping, consider the mapping from type "S: set of T" to type "L: list of T", where the element type T of S and L is integer. Suppose an expression "{1, 2, 3} SUBSET v_s" occurs in the upper level. S is not compatible with L, so the proper subset operator (SUBSET) must be mapped.

```
IMPL(SUBSET(s1: S, s2: S)) ==
        list_len(IMPL(s1)) ≠ list_len(IMPL(s2))
    &   FORALL i: integer
            (   1 ≤ i
            &   i ≤ list_len(IMPL(s1))
        →       EXISTS j: integer
                (   1 ≤ j
                &   j ≤ list_len(IMPL(s2))
                &   IMPL(s2)[j] = IMPL(s1)[i]))
```

In this case, the implementation of subset is defined such that whenever s1 is a proper subset of s2 in the upper level, the lists corresponding to s1 and s2 in the lower level do not have the same length and every element that is on the list IMPL(s1) is on the list IMPL(s2).

There are several things to note about this mapping. First, IMPL is allowed to be recursive on the structure of the parse tree. That is, for an operator $op(p_1, ..., p_n)$, IMPL(op(...)) is allowed to reference $IMPL(p_1), ..., IMPL(p_n)$. This allows the operator mappings to be significantly simplified because it is not necessary to describe how each operand is mapped. The operand mappings are described individually in their own mappings that can be reused in each operator mapping. For example, in the mapping for ISIN, it is not necessary to describe how a set of type S is translated into a list of type L. This is described by a separate mapping. As a consequence of allowing recursion, the translation of an upper level expression cannot simply traverse the parse tree of the expression and replace each mapped object by its right hand side. Instead, the replacement algorithm is directed by the IMPL mapping. That is, the replacement algorithm must call itself whenever IMPL is used in the right hand side of a mapping that is currently being used for replacement.

The other thing to note is that operators may take items other than variables as operands. When a variable v is given as an operand, IMPL(v) is well-defined since all variables must be mapped. The operators may also take explicitly valued constants (e.g. 5, {3, 6}, etc.) and imported variables as operands. This means that an IMPL mapping must be defined to map these types of operands to an equivalent lower level expression of the correct type. In the above example, IMPL(s1) is referenced in the definition of SUBSET and the set {1, 2, 3} is used as an operand to SUBSET in the upper level, so IMPL must define how the set {1, 2, 3} is mapped to LISTDEF (1, 2, 3) and in general how any constant or imported set of integers is mapped to a list of integers. Like operators, a natural constant mapping is defined as follows.

    IMPL_0(c: T) == c for any type T that has a built-in supertype (i.e. integer, real, boolean, or id)

List and set constants are mapped using the natural operator mapping.

```
IMPL_0(LISTDEF(e_1, ..., e_n)) ==
    LISTDEF(IMPL(e_1), ..., IMPL(e_n))
IMPL_0({e_1, ..., e_n}) ==
    {IMPL(e_1), ..., IMPL(e_n)}
IMPL_0(SETDEF e: T (P(e))) ==
    SETDEF e: IMPL(T) (IMPL(P(e)))
```

In these mappings, the values of a built-in type are mapped to the same values, a list of elements is mapped to a list of the implementation of each element, and a set of elements is mapped to a set of the implementation of each element. For each operator mapping $IMPL(op(p_1, ..., p_i: T, ..., p_n))$ that references $IMPL(p_i)$ such that IMPL(c: T) has not been defined and $IMPL\_0(p_i)$ is either undefined or causes a type mismatch when exchanged for $IMPL(p_i)$, the user must define a mapping IMPL(c: T). If no such mapping is required, IMPL(c: T) is defined to be IMPL_0(c: T).

In general, an element of type T in the upper level may be mapped to more than one value of type IMPL(T) at the lower level. For example, consider the mapping from type S to type L. In this mapping, a set v_s in the upper level maps to a list v_l in the lower level such that v_l contains exactly the elements that are in v_s. Lists, however, are ordered, so the elements in v_s may occur in v_l in any order. Therefore, v_s maps to |v_s|! different lists in the lower level. In general, it is undesirable to limit the value that can be chosen in the lower level, which in turn would limit implementation possibilities. For example, if type T was totally ordered, v_l could be chosen such that if t1 and t2 were in v_s and t1 < t2, then t1 would occur before t2 in v_l. In some cases, however, it is not possible to choose one particular value at the lower level. If T is an undefined type, there is no way to describe a transformation from v_s to a specific v_l in the ASTRAL base logic because nothing is known about elements of type T.

To facilitate such mappings, the choose operator, **choose e: T (P(e))**, is introduced into the ASTRAL language, which corresponds to Hilbert's ε-operator [Lei 69]. The value of the expression **choose e: T (P(e))** is an element e of type T, such that the ASTRAL predicate P(e) holds if such an element exists. If more than one such element exists, the operator nondeterministically chooses one of those elements. If no such element exists, the operator nondeterministically chooses some element of T.

With the choose operator in the language, defining element transformations becomes much simpler. For example, consider the mapping from elements of type S to elements of type L.

```
IMPL(v_s: S) ==
     choose v_l: L
         (   list_len(v_l) = set_size(IMPL_0(v_s))
         &   FORALL e: IMPL(T)
                 (   e ISIN IMPL_0(v_s)
             ↔       EXISTS i: integer
                         (   1 ≤ i
                         &   i ≤ list_len(v_l)
                         &   v_l[i] = e)))
```

This mapping states that a constant or imported variable v_s of type S is mapped to a list v_l of type L such that the length of v_l is equal to the cardinality of v_s and every element in v_s is on v_l. Note that in this mapping, the natural mapping IMPL_0 is referenced to avoid any reference to an upper level term (in this case v_s) in the right hand side of the mapping. Although it is possible to avoid referencing upper level terms in most cases, it is impossible to avoid this in all cases. In particular, when mapping constants, it is sometimes necessary to choose a replacement expression based on the actual value of the constant in the upper level. Most notably, when mapping enumerated types, it is necessary to reference upper level enumerated constants in the right hand side of the IMPL mapping. For example, consider an upper level enumerated type **gate_u: (open, closed)** that is mapped to a lower level type **gate_l: (open, closed, opening, closing)** such that **closed** maps to **closed** and **open** maps to one of **open, opening,** or **closing**. In this mapping, there is no way to map an arbitrary constant of type **gate_u** to a constant of type **gate_l** without selecting a value based on **gate_u**. To accommodate such mappings, a single case split is allowed on the upper level constant that is being mapped such that each case corresponds to an explicit constant value. For example, arbitrary **gate_u** constants can be mapped as follows:

```
IMPL(c: gate_u) ==
     CASE c OF
         open:
             choose e: gate_l
                 (   e = open
                 |   e = opening
                 |   e = closing)
         closed:
             closed
     ESAC
```

When c is an actual constant value, the IMPL replacement algorithm uses the case information to choose the correct replacement. When c is an imported variable, the right side of the mapping is substituted as is, which is well-defined since the upper level type must be globally defined and the interface to the process must stay fixed from the top level, so types available at the top level are still available at the lower levels.

## 6. The Mult_Add Circuit

The specification of the refinement of the Mult_Add circuit in Figure 3 is shown below using the new parallel refinement mechanisms. Each multiplier has a single exported transition multiply, which computes the product of two inputs. The adder has a single transition add, which computes the sum of the two multiplier outputs.

```
PROCESS Multiplier                      PROCESS Adder
EXPORT                                  IMPORT
    multiply, product                       M1, M1.product, M1.multiply,
VARIABLE                                    M2, M2.product, M2.multiply
    product: integer                    EXPORT    sum
TRANSITION multiply(a, b: integer)      VARIABLE  sum: integer
    ENTRY   [TIME: 2]                   TRANSITION add
        EXISTS t: time                      ENTRY   [TIME: 1]
            (End(multiply, t))                  M1.End(multiply, now)
        → now - End(multiply) ≥ 1           & M2.End(multiply, now)
    EXIT                                    EXIT
        product = a * b                         sum = M1.product + M2.product
```

The lower level consists of two instances of the Multiplier process type and one instance of the Adder process type.

```
PROCESSES
    M1, M2: Multiplier
    A1: Adder
```

The output variable of the upper level process is mapped to the sum variable of the adder.

    IMPL(output) == A1.sum

The duration of the compute transition is the sum of the multiply transition and the add transition in the lower level.

    IMPL(dur1) == 3

When compute starts in the upper level, multiply starts on both M1 and M2. When compute ends in the upper level, add ends on A1. When compute is called in the upper level with inputs a, b, c, and d, multiply is called on M1 with inputs a and b and multiply is called on M2 with inputs c and d.

```
IMPL(Start(compute, now)) ==           IMPL(End(compute, now)) == A1.End(add, now)
    M1.Start(multiply, now)            IMPL(Call(compute(a, b, c, d), now)) ==
  & M2.Start(multiply, now)                M1.Call(multiply(a, b), now)
                                         & M2.Call(multiply(c, d), now)
```

# 7. Proof Obligations for Parallel Refinement Mechanisms

The goal of the refinement proof obligations is to show that any properties that hold in the upper level hold in the lower level without actually reproving the upper level properties in the lower level. In order to show this, it must be shown that the lower level correctly implements the upper level. ASTRAL properties are interpreted over execution histories, which are described by the values of state variables and the start, end, and call times of transitions at all times in the past back to the initialization of the system. A lower level correctly implements an upper level if the implementation of the execution history of the upper level is equivalent to the execution history of the lower level. This corresponds to proving the following four statements.

(V)  Any time a variable has one of a set S of possible values in the upper level, the implementation of the variable has one of a subset of the implementation of S in the lower level.

(C)  Any time the implementation of a variable changes in the lower level, a transition ends in the upper level.

(S)  Any time a transition starts in the upper level, the implementation of the transition starts in the lower level and vice-versa.

(E)  Any time a transition ends in the upper level, the implementation of the transition ends in the lower level and vice-versa.

If these four items can be shown, then any property that holds in the upper level is preserved in the lower level because the structures over which the properties are interpreted are identical over the implementation mapping.

## 7.1. Proof Obligations

Instead of proving directly that the mappings hold at all times, it can be shown that the mappings hold indirectly by proving that they preserve the axiomatization of the ASTRAL abstract machine; thus they preserve any reasoning performed in the upper level. This can be accomplished by proving the implementation of each abstract machine axiom.

To perform the proofs, the following assumption must be made about calls to transitions in each lower level process.

```
impl_call: ASSUMPTION
    FORALL tr_ll: trans_ll, t1: time
      (   Exported(tr_ll)
       &  past(Call(tr_ll, t1), t1)
    →   EXISTS tr_ul: trans_ul
          (   FORALL t2: time
                (   IMPL(past(Call(tr_ul, t2), t2))
                 →    past(Call(tr_ll, t2), t2))
           &  IMPL(past(Call(tr_ul, t1), t1))))
```

This assumption states that any time a lower level exported transition is called, there is some call mapping that references a call to the transition that holds at the same time. This means that if one transition of a "conjunctive" mapping is called, then all transitions of the mapping are called. That is, it is not possible for a lower level transition to be called such that the call mapping for some upper level transition does not hold. For example, consider the mapping for the compute transition of the Mult_Add circuit.

    IMPL(Call(compute(a, b, c, d), now)) ==
        M1.Call(multiply(a, b), now)
      & M2.Call(multiply(c, d), now)

In this case, impl_call states that any time multiply is called on M1, multiply is called on M2 at the same time and vice-versa.

An assumption is also needed to assure that whenever the parameters of an upper level exported transition are distributed among multiple transitions at the lower level, the collection of parameters for which the lower level transitions execute come from a single set of call parameters. For example, in the Mult_Add circuit, the compute transition in the upper level may be called with two sets of parameters {1, 2, 3, 4} and {5, 6, 7, 8} at the same instant. In the lower level implementation, the multiply transition of each multiplier takes two of the parameters from each upper level call. Thus, in the example, multiply is enabled on M1 for {1, 2} and {5, 6} and on M2 for {3, 4} and {7, 8}. Without an appropriate assumption, M1 may choose {1, 2} and M2 may choose {7, 8}, thus computing the product for {1, 2, 7, 8}, which was not requested at the upper level.

The implementation of the call_fire_parms axiom provides the appropriate restriction. The *impl_call_fire_parms* assumption states that any time the mapped start of an exported parameterized transition occurs, the mapped parameters for which the mapped transition fired came from the set of mapped call parameters that have not yet been serviced at that time.

    impl_call_fire_parms: ASSUMPTION
      FORALL tr1: transition, t3: time
        (   Exported(tr1)
          & Has_Parms(tr1)
          & IMPL(past(Start(tr1, t3), t3))
      →     EXISTS t1: time
          (   t1 ≤ t3
          & IMPL(past(Call(tr1, t1), t1))
          & IMPL(Fire_Parms(tr1, t3)) ISIN IMPL(Call_Parms(tr1, t1))
          & FORALL t2: time
             (   t1 ≤ t2 & t2 < t3
             & IMPL(past(Start(tr1, t2), t2))
            →   IMPL(Fire_Parms(tr1, t2)) ≠ IMPL(Fire_Parms(tr1, t3)))))

Note that impl_call_fire_parms must be stated as an assumption and not as an obligation because there is no way to specify the lower level processes such that they will collectively make the same nondeterministic choice of which set of call parameters to service. Since this behavior cannot be specified, impl_call_fire_parms cannot be proved as an obligation. The other portions of the impl_call_fire_parms assumption hold by the restriction on call mappings for exported parameterized transitions as mentioned in Section 5.2.

In the axiomatization of the ASTRAL abstract machine [Kol 99], the predicate "Fired(tr1, t1)" is used to denote that the transition tr1 fired at time t1. If Fired(tr1, t1) holds, then it is derivable that a start of tr1 occurred at t1 and an end of tr1 occurred at t1 + Duration(tr1). Additionally, since an end of tr1 can only occur at t1 when Fired(tr1, t1 - Duration(tr1)) holds and the time parameter of Fired is restricted to be nonnegative, it is known that an end can only occur at times greater than or equal to the duration of the transition. In the parallel refinement mechanisms, the start and end of upper level transitions are mapped by the user, so it is unknown whether these properties of end still hold. Since the axioms rely on these properties, they must be proved explicitly as proof obligations. The *impl_end1* obligation ensures that the mapped end of a transition can only occur after the mapped duration of the transition has elapsed.

```
impl_end1: OBLIGATION
    FORALL tr1: transition, t1: time
        (   IMPL(past(End(tr1, t1), t1))
    →       t1 ≥ IMPL(Duration(tr1)))
```

The *impl_end2* obligation ensures that for every mapped start of a transition, there is a corresponding mapped end of the transition, that for every mapped end, there is a corresponding mapped start, and that mapped starts and mapped ends are separated by the mapped duration of the transition.

```
impl_end2: OBLIGATION
    FORALL tr1: transition, t1: time, t2: time
        (   t1 = t2 - IMPL(Duration(tr1))
    →       (   IMPL(past(Start(tr1, t1), t1))
        ↔       IMPL(past(End(tr1, t2), t2))))
```

The following obligations are the mappings of the ASTRAL abstract machine axioms except for call_fire_parms, which is discussed above. The *impl_trans_entry* obligation ensures that any time the mapped start of a transition occurs, the mapped entry assertion of the transition holds.

```
impl_trans_entry: OBLIGATION
    FORALL tr1: transition, t1: time
        (   IMPL(past(Start(tr1, t1), t1))
    →       IMPL(Entry(tr1, t1)))
```

The *impl_trans_exit* obligation ensures that any time the mapped end of a transition occurs, the mapped exit assertion of the transition holds.

```
impl_trans_exit: OBLIGATION
    FORALL tr1: transition, t1: time
        (   IMPL(past(End(tr1, t1), t1))
    →       IMPL(Exit(tr1, t1)))
```

The *impl_trans_called* obligation ensures that any time the mapped start of an exported transition occurs, a mapped call has been issued to the transition but not yet serviced.

```
impl_trans_called: OBLIGATION
    FORALL tr1: transition, t1: time
        (   IMPL(past(Start(tr1, t1), t1))
    &       Exported(tr1)
    →       IMPL(Issued_Call(tr1, t1)))
```

The *impl_trans_mutex* obligation ensures that any time the mapped start of a transition occurs, no other mapped start of a transition can occur until the mapped duration of the transition has elapsed.

```
impl_trans_mutex: OBLIGATION
    FORALL tr1: transition, t1: time
        (   IMPL(past(Start(tr1, t1), t1))
    →       FORALL tr2: transition
                (   tr2 ≠ tr1
    →               ~IMPL(past(Start(tr2, t1), t1)))
        &   FORALL tr2: transition, t2: time
                (   t1 < t2 & t2 < t1 + IMPL(Duration(tr1))
    →               ~IMPL(past(Start(tr2, t2), t2))))
```

The *impl_trans_fire* obligation ensures that any time the mapped entry assertion of a transition holds, a mapped call has been issued to the transition but not yet serviced if the transition is exported, and no mapped start of a transition has occurred within its mapped duration of the given time, a mapped start will occur.

```
impl_trans_fire: OBLIGATION
    FORALL t1: time
        (   EXISTS tr1: transition (IMPL(Enabled(tr1, t1)))
        &   FORALL tr2: transition, t2: time
                (   t1 - IMPL(Duration(tr2)) < t2 & t2 < t1
    →               ~IMPL(past(Start(tr2, t2), t2)))
    →       EXISTS tr1: transition (IMPL(past(Start(tr1, t1), t1))))
```

The *impl_vars_no_change* obligation ensures that mapped variables only change value when the mapped end of a transition occurs.

```
impl_vars_no_change: OBLIGATION
    FORALL t1: time, t3: time
        (   t1 ≤ t3
        &   FORALL tr2: transition, t2: time
                (   t1 < t2 & t2 ≤ t3
    →               ~IMPL(past(End(tr2, t2), t2)))
    →       FORALL t2: time
                (   t1 ≤ t2 & t2 ≤ t3
    →               IMPL(Vars_No_Change(t1, t2))))
```

The *impl_initial_state* obligation ensures that the mapped initial clause holds at time zero.

```
impl_initial_state: OBLIGATION
    IMPL(Initial(0))
```

Besides the abstract machine axioms, the local proofs of ASTRAL process specifications can also reference the local axiom clause of the process (which is empty in the Mult_Add circuit). Since this clause can be used in proofs and the constants referenced in the clause can be implemented at the lower level, the mapping of the local axiom clause of the upper level must be proved as a proof obligation. The *impl_local_axiom* obligation ensures that the mapped axiom clause holds at all times. In order to prove this obligation, it may be necessary to specify local axioms in the lower level processes that satisfy the implementation of the upper level axiom clause.

```
impl_local_axiom: OBLIGATION
    FORALL t1:time (IMPL(Axiom(t1)))
```

To prove the refinement proof obligations, the abstract machine axioms can be used in each lower level process. For example, to prove the impl_initial_state obligation, the initial_state axiom of each lower level process can be asserted.

**7.2. Correctness of Proof Obligations**

The proof obligations for the parallel refinement mechanisms as stated above are sufficient to show that for any invariant I that holds in the upper level, IMPL(I) holds in the lower level. Consider the correctness criteria (V), (C), (S), and (E) above. (V) is satisfied because by impl_initial_state, the values of the implementation of the variables in the lower level must be consistent with the values in the upper level. Variables in the upper level only change when a transition ends and at these times, the implementation of the variables in the lower level change consistently by impl_trans_exit. (C) is satisfied because the implementation of the variables in the lower level can only change value when the implementation of a transition ends by impl_vars_no_change. The forward direction of (S) is satisfied because whenever an upper level transition fires, a lower level transition will fire by impl_trans_fire. The reverse direction of (S) is satisfied because whenever the implementation of a transition fires in the lower level, its entry assertion holds by impl_trans_entry, it has been called by impl_trans_called, and no other transition is in the middle of execution by impl_trans_mutex. (E) is satisfied because (S) is satisfied and by impl_end1 and impl_end2, any time a start occurs, a corresponding end occurs and vice-versa.

More formally, any time an invariant I can be derived in the upper level, it is derived by a sequence of transformations from I to TRUE, $I \vdash_{f1/a1} I_1 \vdash_{f2/a2} ... \vdash_{fn/an} \text{TRUE}$, where each transformation $f_i/a_i$ corresponds to the application of a series $f_i$ of first-order logic axioms and a single abstract machine axiom $a_i$. Since the implementation of each axiom of the ASTRAL abstract machine is preserved by the parallel refinement proof obligations, a corresponding proof at the lower level $\text{IMPL}(I) \vdash_{f1'/impl\_a1} \text{IMPL}(I_1) \vdash_{f2'/impl\_a2} ... \vdash_{fn'/impl\_an} \text{TRUE}$ can be constructed by replacing the application of each abstract machine axiom $a_i$ by impl_$a_i$. Additionally, each series $f_i$ of first-order logic axioms is replaced by a series $f_i'$ that takes any changes to the types of variables and constants into consideration.

## 8. Proof of Mult_Add Circuit Refinement

This section shows the application of the parallel refinement proof obligations to the Mult_Add circuit. The obligations below were obtained from the corresponding obligations of the previous section by expanding the IMPL mapping appropriately, replacing quantification over transitions with the actual transitions of the Mult_Add circuit, and performing some minor simplifications.

### 8.1. Impl_end1 Obligation

```
FORALL t1: time
    (   past(A1.End(add, t1), t1)
 →      t1 ≥ 3)
```

By the entry assertion of add, multiply must end when add starts. The duration of add is 1 and the duration of multiply is 2, so the earliest add can end is at time 3. Thus, impl_end1 holds.

### 8.2. Impl_end2 Obligation

>   FORALL t1: time  
>    ( past(M1.Start(multiply, t1 - 3), t1 - 3)  
>    & past(M2.Start(multiply, t1 - 3), t1 - 3)  
>   ↔ past(A1.End(add, t1), t1))

For forward direction, it must be shown that add ends on A1 at t1; thus, it starts at t1 - 1. From the antecedent, multiply ends on both M1 and M2 at t1 - 1 so the entry assertion of add holds on A1 at time t1 - 1. A1 must be idle or else from the entry of add, multiply ended in the interval (t1 - 2, t1 - 1), which is not possible since multiply was still executing on M1 and M2 in that interval. Therefore, add starts at t1 - 1 on A1; thus, it ends at t1.

For the reverse direction, add starts on A1 at t1 - 1 from the antecedent. From the entry of add, multiply ends on both M1 and M2 at t1 - 1; so, it starts at t1 - 3. Thus, the reverse direction holds and impl_end2 holds.

### 8.3. Impl_trans_entry Obligation

>   FORALL t1: time  
>    ( past(M1.Start(multiply, t1), t1)  
>    & past(M2.Start(multiply, t1), t1)  
>   → TRUE)

This formula holds trivially.

### 8.4. Impl_trans_exit Obligation

>   FORALL t1: time  
>    ( past(A1.End(add, t1), t1)  
>   → FORALL a, b, c, d: integer  
>     ( past(M1.Start(multiply(a, b), t1 - 3), t1 - 3)  
>     & past(M2.Start(multiply(c, d), t1 - 3), t1 - 3)  
>     → past(A1.sum, t1) = a * b + c * d))

By the exit assertion of add, past(A1.sum, t1) = past(M1.product, t1 - 1) + past(M2.product, t1 - 1). From the entry of add, multiply ends on both M1 and M2 at t1 - 1. Since multiply ends on M1 and M2 at t1 - 1, it starts on M1 and M2 at t1 - 3 for two pairs of parameters (a, b) and (c, d), respectively, which were provided by the external environment. By the exit assertion of multiply, past(M1.product, t1 - 1) = a * b and past(M2.product, t1 - 1) = c * d, so past(A1.sum, t1) = a * b + c * d. Thus, impl_trans_exit holds.

### 8.5. Impl_trans_called Obligation

    FORALL t1: time
        (   past(M1.Start(multiply, t1), t1)
        &   past(M2.Start(multiply, t1), t1)
    →       EXISTS t2: time
            (   t2 ≤ t1
            &   past(M1.Call(multiply, t2), t1)
            &   past(M2.Call(multiply, t2), t1)
            &   FORALL t3: time
                (   t2 ≤ t3 & t3 < t1
                →   ~ (   past(M1.Start(multiply, t3), t3)
                      &   past(M2.Start(multiply, t3), t3)))))

Since multiply started on M1 (M2) at time t1, by trans_called applied on process M1 (M2), multiply was called at some time t2 ≤ t1 and multiply has not started on M1 (M2) in the interval [t2, t1). By impl_call, the time that multiply was called on M1 and M2 must be the same. Thus, impl_trans_called holds.

### 8.6. Impl_trans_mutex Obligation

    FORALL t1: time
        (   past(M1.Start(multiply, t1), t1)
        &   past(M2.Start(multiply, t1), t1))
    →   FORALL t2: time
            (   t1 < t2 & t2 < t1 + 3
            →   ~ (   past(M1.Start(multiply, t2), t2)
                  &   past(M2.Start(multiply, t2), t2))))

Since multiply started on M1 (M2) at time t1, by trans_mutex applied on process M1 (M2), nothing can fire on M1 (M2) until time t1 + 2. The multiply transition, however, is the only transition of M1 (M2) and multiply is not enabled until 1 time unit after the end of the last multiply; so, it cannot start until t1 + 3. Thus, impl_trans_mutex holds.

### 8.7. Impl_trans_fire Obligation

    FORALL t1: time
        (   EXISTS t2: time
            (   t2 ≤ t1
            &   past(M1.Call(multiply, t2), t1)
            &   past(M2.Call(multiply, t2), t1)
            &   FORALL t3: time
                (   t2 ≤ t3 & t3 < t1
                →   ~ (   past(M1.Start(multiply, t3), t3)
                      &   past(M2.Start(multiply, t3), t3))))
        &   FORALL t2: time
            (   t1 - 3 < t2 & t2 < t1
            →   ~ (   past(M1.Start(multiply, t2), t2)
                  &   past(M2.Start(multiply, t2), t2)))
    →   past(M1.Start(multiply, t1), t1)
    &   past(M2.Start(multiply, t1), t1))

To prove this obligation, it is first necessary to prove that M1.Start(multiply) and M2.Start(multiply) always occur at the same time. This can be proved inductively. At time zero, both M1 and M2 are idle. By impl_call, if multiply is called on either M1 or M2, multiply is called on both M1 and M2. At time zero, multiply cannot have ended, thus the entry assertion of multiply is true, so if both are called, both fire. If neither is called, then neither can fire. For the inductive case, assume M1.Start(multiply) and M2.Start(multiply) have occurred at the same time up until time T0. Suppose multiply occurs on M1 (the M2 case is similar), then M1 was idle, multiply has been called since the last start, and it has been at least one time unit since multiply ended on M1. M2 cannot be executing multiply at T0 or else M1 must also be executing multiply by the inductive hypothesis, thus M2 must be idle. Similarly, it must have been at least one time unit since multiply ended on M2. By impl_call, multiply must have been called on M2 since it was called on M1. Thus, multiply is enabled on M2; so, it must fire. Therefore, M1.Start(multiply) and M2.Start(multiply) always occur at the same time. From this fact

> FORALL t3: time
>     ( t2 ≤ t3 & t3 < t1
> →    ~ ( past(M1.Start(multiply, t3), t3)
>         & past(M2.Start(multiply, t3), t3)))

is equivalent to

> FORALL t3: time
>     ( t2 ≤ t3 & t3 < t1
> →    ~past(M1.Start(multiply, t3), t3)
>      & ~past(M2.Start(multiply, t3), t3))

Since nothing has started in the interval (t1 - 3, t1), nothing can end in the interval (t1 - 1, t1 + 2), thus the entry assertion of multiply on M1 is satisfied at t1. Since the entry of multiply holds, multiply has been called but not yet serviced, and M1 is idle, multiply starts on M1 at t1. Since multiply always starts on both M1 and M2 at the same time as shown above, impl_trans_fire holds.

## 8.8. Impl_vars_no_change Obligation

> FORALL t1, t3: time
>     ( t1 ≤ t3
>     & FORALL t2: time
>         ( t1 < t2 & t2 ≤ t3
>         →    ~past(A1.End(add, t2), t2))
> →    FORALL t2: time
>         ( t1 ≤ t2 & t2 ≤ t3
>         →    past(A1.sum, t1) = past(A1.sum, t2)))

This formula holds by the vars_no_change axiom applied on process A1.

### 8.9. Results of Proof Obligations

The impl_initial_state and impl_local_axiom obligations trivially hold, because the initial and axiom clauses are both "TRUE". Since the proof obligations hold for the Mult_Add circuit, the lower level is a correct refinement of the upper level and thus the implementation of the upper level invariant, shown below, holds in the lower level.

$$
\begin{aligned}
& \text{FORALL t1: time, a, b, c, d: integer} \\
& \quad (\quad \text{M1.Start(multiply(a, b), t1 - 3)} \\
& \quad \&\quad \text{M2.Start(multiply(c, d), t1 - 3)} \\
& \rightarrow \quad \text{FORALL t2: time} \\
& \qquad (\quad t1 + dur1 \leq t2 \\
& \qquad \&\quad t2 \leq now \\
& \rightarrow \quad \text{past(A1.sum, t2)} = a * b + c * d))
\end{aligned}
$$

## 9. Parallel Phone System

The previous example has shown that the parallel refinement mechanisms can express the parallel implementation of a simple system in a simple and straightforward manner. Furthermore, the proof obligations for a simple implementation were themselves simple. In this section, the refinement of a much more complex system will be discussed along with the application of the proof obligations to it. From the following example, it will be shown that the parallel refinement mechanisms can be used to express very complex parallel implementations, but at a cost of complicating the proofs of the proof obligations. In order to simplify the presentation of the refinement of such a complex system, we focus on a few significant portions of it and postpone several fine-grain details and proofs until the appendix. The system considered here is a slightly modified version of the phone system defined in [CGK 97]. It consists of a set of phones that need various services (e.g. getting a dial tone, processing digits entered into the phone, and making a connection to the requested phone) as well as a set of central control centers that perform the services. Each control center is responsible for the phones belonging to its area, and it is provided with all the functionality needed to set up a local call. Control centers are also intended to deal with long distance calls (i.e. calls to other areas). Calls to outside areas are modeled by exported variables (i.e. the data is sent to the external environment), while calls from an outside area are modeled as exported transitions (i.e. the parameters of calls to exported transitions from the external environment provide the information associated with each outside call). The example is a simplification of a real phone system. Every local phone number is seven digits long, area codes are three digits long, a customer can be connected to at most one other phone (either local or in another area), and ongoing calls cannot be interrupted. The main requirement of the system is that phones will be given a dial tone within two seconds.

The specification of the central control, which is the core of the whole system, is articulated in three layers. The goal of the top level is to provide an abstract and global view of the supplied services in such a way that the user can have a complete and precise knowledge of the external behavior, both in terms of functions performed and in terms of service times of the central control, but the designer still has total freedom for implementation policies. In fact, as a result, the description provided in [CGK 97] is just an alternative implementation of the top level description given below, which differs from this version in that services are granted sequentially rather than in parallel.

## 9.1. Top Level of the Central Control

To achieve this goal (i.e. to allow the implementation of services both asynchronously in parallel and strictly sequentially, as suggested by Figures 5 and 6), the top level is specified such that a *set* of services can start and a *set* of services can end at every time unit in the system (for simplicity, discrete time is assumed). In these figures, Ti_si.Pk denotes providing service si to phone k.

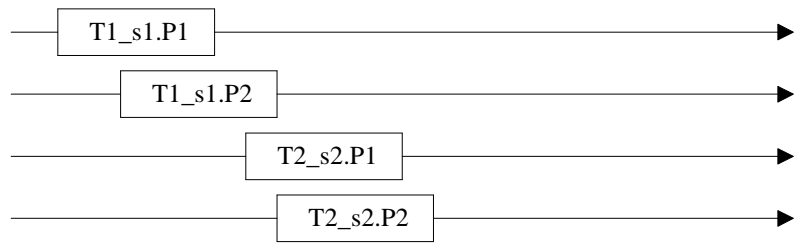

Figure 5: Parallel implementation of different services on several phones

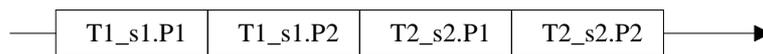

Figure 6: Sequential implementation of different services on several phones

The service of a phone is split into the beginning of servicing and the completion of servicing through two transitions: Begin_Serve and Complete_Serve. In other words, instead of assigning ASTRAL transitions to single services or groups thereof, we only make the beginning and the end of services visible at the top level. In this way, we do not commit too early to a fixed duration of the implementation of the service, stating only when a service will begin and when it will be completed. Thus, the durations of Begin_Serve and Complete_Serve are set to a constant serve_dur, where 2 * serve_dur is chosen to be a divisor of the duration of every service.

Begin_Serve and Complete_Serve execute cyclically and indicate the start and end, respectively, of the execution of several different functions, each of which corresponds to a transition of the [CGK 97] central control (excluding the part referring to long distance calls). The functions executed by Begin_Serve and Complete_Serve are:

|        |                    |         |                      |
|--------|--------------------|---------|----------------------|
| (GDT)  | Give_Dial_Tone     | (ERB)   | Enable_Ringback      |
| (PD)   | Process_Digit      | (DRBP)  | Disable_Ringback_Pulse |
| (PC)   | Process_Call       | (ST)    | Start_Talk           |
| (ER)   | Enable_Ring        | (TC)    | Terminate_Connection |
| (DRP)  | Disable_Ring_Pulse | (GA)    | Generate_Alarm       |

For each function g, let Dur_g and Exit_g refer to the duration and the exit assertion, respectively, of the [CGK 97] transition corresponding to function g. The main functions used for illustration in the remainder of the paper are GDT, PD, and PC. The corresponding transitions of [CGK 97] are specified below. Note that the portion of each transition referring to long distance calls was removed and that the exit assertions of Give_Dial_Tone and Process_Digit were rewritten slightly (without changing the effective behavior of the system) to keep changes to the Number variable exclusive to Process_Digit.

```
Give_Dial_Tone(P: phone)
    ENTRY     [TIME:    Tim1]
        P.Offhook
      & Phone_State(P) = Idle
    EXIT
        Phone_State(P) BECOMES Ready_To_Dial

Process_Digit(P: phone)
    ENTRY     [TIME:    Tim2]
        P.Offhook
      & (  Phone_State(P) = Ready_To_Dial
        |  Phone_State(P) = Dialing
        &  Count(P) < 7
        &  P.End(Enter_Digit) > Change(Count(P)))
    EXIT
        IF Phone_State'(P) = Ready_To_Dial
        THEN
              Number(P) BECOMES LISTDEF(P.Next_Digit')
           &  Phone_State(P) BECOMES Dialing
        ELSE
              Number(P) BECOMES Number'(P) CONCAT LISTDEF(P.Next_Digit')
        FI

Process_Call(P: phone)
    ENTRY     [TIME:    Tim3]
        P.Offhook
      & Count(P) = 7
      & Phone_State(P) = Dialing
      & ~Get_ID(Number(P)).Offhook
      & Phone_State(Get_ID(Number(P))) = Idle
    EXIT
        Phone_State(Get_ID(Number'(P))) = Ringing
      & Phone_State(P) = Waiting
      & Connected_To(P) = Get_ID(Number'(P))
      & Connected_To(Get_ID(Number'(P))) = P
      & FORALL P1: phone
            (  P1 ≠ P & P1 ≠ Get_ID(Number'(P))
            →  NOCHANGE(Phone_State(P1))
            &  NOCHANGE(Connected_To(P1)))
```

All functions are specified similarly to those in [CGK 97], but instead of specifying a separate transition for each, the entry assertion of **Begin_Serve** and the exit assertion of **Complete_Serve** are the conjunctions of the entry assertions of each function and the exit assertions of each function, respectively. In addition, each function is specified to service a set of phone processes instead of a single phone. A parameterized variable **serving(P)** records when each phone P is being served. When **serving(P)** changes to true for a phone P at time t, P began being served at t - serve_dur. Thus, when the duration of the function that was serving the phone elapses from this time, **Complete_Serve** carries out the effect of the function on the phone's state and resets **serving** for that phone to false.

For each function g above, a set of phones W_g is defined, which is the set of phones waiting to be serviced by the function g. These sets are described as follows.

```
(W_GDT)    setdef P: phone (  P.Offhook & Phone_State(P) = Idle)
(W_PD)     setdef P: phone (  P.Offhook
                           & (  Phone_State(P) = Ready_To_Dial
                              | Phone_State(P) = Dialing
                              & Count(P) < 7
                              & P.End(Enter_Digit) > Change(Count(P))))
(W_PC)     setdef P: phone (  P.Offhook & Count(P) = 7
                           & Phone_State(P) = Dialing
                           & ~Get_ID(Number(P)).Offhook
                           & Phone_State(Get_ID(Number(P))) = Idle)
...
```

To allow both a sequential and a parallel implementation, it is necessary for the top level specification to allow the possibility of multiple actions occurring at the same time without actually requiring multiple actions to occur. This is achieved by limiting the number of phones that can be serviced at any given time to be less than a constant K_max. In the sequential refinement, K_max is mapped to one, indicating that only one phone at a time can be serviced. In the parallel refinement, K_max is mapped to the sum of the capacities of the individual servers, indicating that as many phones as it is possible for the servers to serve can be serviced in parallel. In general, for each function g, let the set W_g be the set of phones that satisfy the entry assertion of the transition associated with g. Let K_W_g be the maximum number of phones that can be served by the function g at any time and K_max be the maximum number of phones that can be served by any function at any time. In the following definitions of Begin_Serve and Complete_Serve, quantification over the functions g of the central control is used to simplify the presentation. The quantifiers can be expanded out over the 10 functions of the central control. For each function g, let serving_g be defined as setdef P: phone (serving(P) & P ISIN past(W_g, change(serving(P)) - serve_dur). That is, serving_g specifies the set of phones currently being served by g. This definition is necessary since each W_g changes dynamically over time according to the behavior of the phone processes. Additionally, let serving_all be defined as setdef P: phone (serving(P)), which is the set of all phones being served.

Begin_Serve(S) is enabled for a nonempty set S when

- now is a multiple of 2 * serve_dur
- S is the union of the sets S_g, where for each function g,
    1. S_g only contains phones that are in W_g
    2. S_g does not contain any phones currently being served
    3. S_g $\cup$ serving_g contains at most K_W_g phones
- S $\cup$ serving_all contains at most K_max phones
- if S $\cup$ serving_all contains less than K_max phones, then for each function g, either S_g $\cup$ serving_g is at maximum capacity or all the phones of W_g are in S $\cup$ serving_all

The exit assertion of Begin_Serve specifies that the set of phones that begin being served is equal to the set parameter S.

```
Begin_Serve(S: nonempty_set_of_phone)
    ENTRY   [TIME:  serve_dur]
        now MOD (2 * serve_dur) = 0
      & EXISTS S_GDT, S_PD, ..., S_TC: set_of_phone
            (   S_GDT CONTAINED_IN W_GDT
             & S_GDT SET_DIFF serving_all = S_GDT
             & set_size(S_GDT UNION serving_GDT) ≤ K_W_GDT
             & ...
             & S = S_GDT UNION S_PD UNION ... UNION S_TC
             & set_size(S UNION serving_all) ≤ K_max
             & (  set_size(S UNION serving_all) < K_max
               → FORALL g: central function
                     (  set_size(S_g UNION serving_g) = K_W_g
                      | W_g CONTAINED_IN S_g UNION serving_all)))
    EXIT
        FORALL P: phone
            (IF P ISIN S
             THEN serving(P)
             ELSE serving(P) ↔ serving'(P)
             FI)
```

Complete_Serve is enabled when

- now + serve_dur is a multiple of 2 * serve_dur
- enough time has elapsed (i.e. the duration of the function) since some phone began being served such that service for that phone can be completed

The exit assertion of Complete_Serve specifies that all phones that have been served for at least the duration of the appropriate function will complete being served with the variables for each phone changed according to the exit assertion of the appropriate function.

```
Complete_Serve
    ENTRY   [TIME:  serve_dur]
        now MOD (2 * serve_dur) = serve_dur
      & EXISTS P: phone, g: central function
            (  P ISIN serving_g
             & now - change(serving(P)) + serve_dur ≥ Dur_g - serve_dur)
    EXIT
        FORALL P: phone, g: central function
            (IF   P ISIN past(serving_g, now - serve_dur)
              &  now - past(change(serving(P)), now - serve_dur) ≥
                        Dur_g - serve_dur
             THEN
                    Exit_g(P)
                & ~serving(P)
             ELSE
                    serving(P) ↔ serving'(P)
             FI)
```

Notice that the above specifications automatically define the updating of the set of phones. That is, each set W_g is updated according to changes in external processes (e.g. phones becoming offhook) and according to the changes made by the exit assertion of Complete_Serve.

**9.2. Parallel Refinement of Top Level Central Control**

A sequential implementation of the top level central control has been developed and proved correct according to the strategies discussed in Sections 3 and 4. This implementation is presented in Appendix A. Let us now look at a parallel implementation of the same top level central control. As mentioned earlier, this is achieved through two refined layers. In the first refinement, discussed in this section, the central control is split into several parallel server processes, each of which is devoted to a single service of the top level central control. Thus, there is a server devoted to giving dial tone, a server devoted to processing entered digits, and so on. In the second refinement, which will be discussed in Section 9.4, each server of the first refinement is implemented by a parallel array of *microservers*, where each microserver is devoted to providing a single service to a single phone (e.g. processing the call of one particular phone).

The main issue in the first refinement is the mapping of the global state of the central control into disjoint components to be assigned to the different lower level parallel processes. That is, the Phone_State(P) variable in the top level, which holds the state of each phone P (Phone_State can take the values idle, ready_to_dial, ringing, ...), needs to be split among all the servers in the lower level. Figure 7 shows the relationship between the functions and the variables of the central control. A function connected to a variable indicates that the exit assertion of the function sets the variable.

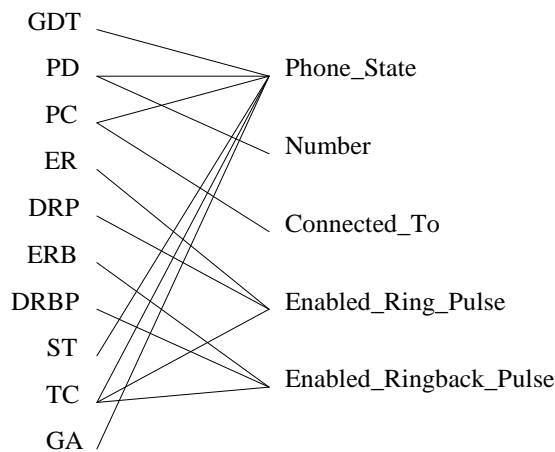

Figure 7: Function and variable relationship in the central control

The most critical variable of the central control is the Phone_State variable, which is set by six of the ten functions. In the following discussion of the parallel refinement of the top level central control, only the implementation of the Phone_State variable and its corresponding type Enabled_State will be described. The other variables can be mapped in a similar fashion.

The type Enabled_State describes all states through which the managing of a call passes during the evolution of the call itself. These states build a sort of "chain" and the various state transformations move the state of each phone call through the chain from Idle to Ready_To_Dial, etc. In the implementation of the top level, each step of this path is executed by a different process. In the ASTRAL model, however, only a single process can change the value of a variable, thus it is not possible to let all of the lower level servers change the same variable directly. A solution to this problem is to map the type Enabled_State into a structure of time fields, where each enumerated constant in Enabled_State has a field in the structure as shown below.

    IMPL(Enabled_State) ==
        STRUCTURE OF (Idle, Ready_To_Dial, Ringing, ...: time)

The basic idea of this mapping is that each field of the structure is a timestamp and the field with the most recent timestamp determines the value of variables of the structure type. Since Enabled_State is not compatible with IMPL(Enabled_State), a mapping for constants must be defined as shown below.

    IMPL(v_es: Enabled_State) ==
        CASE v_es OF
          Idle:
            choose i_v_es: IMPL(Enabled_State)
              (  FORALL f: field (i_v_es[f] = 0)
              |  FORALL f: field
                  (f ≠ Idle → i_v_es[Idle] > i_v_es[f]))
          Ready_To_Dial:
            choose i_v_es: IMPL(Enabled_State)
              (  FORALL f: field
                  (f ≠ Ready_To_Dial → i_v_es[Ready_To_Dial] > i_v_es[f]))
          Ringing:
            choose i_v_es: IMPL(Enabled_State)
              (  FORALL f: field
                  (f ≠ Ringing → i_v_es[Ringing] > i_v_es[f]))
          ...
        ESAC

This mapping states that a constant that is Idle in the upper level maps to a structure of type IMPL(Enabled_State) such that all the fields are zero or the Idle field is greater than all the other fields. For values v other than Idle, a constant maps to a structure such that the field associated with v is greater than all the other fields.

In addition to a mapping for constants of type Enabled_State, a mapping must also be defined for the operators with operands of type Enabled_State. The only operator used on operands of type Enabled_State is the = operator. The mapping for the = operator is shown below.

    IMPL(=(v_es1, v_es2: Enabled_State): bool) ==
        (  (  FORALL f1: field (IMPL(v_es1)[f1] = 0)
          →    FORALL f1: field
              (  IMPL(v_es2)[f1] = 0
              |  (f1 ≠ Idle → IMPL(v_es2)[Idle] > IMPL(v_es2)[f1])))
        |  (  FORALL f1: field (IMPL(v_es2)[f1] = 0)
          →    FORALL f1: field
              (  IMPL(v_es1)[f1] = 0
              |  (f1 ≠ Idle → IMPL(v_es1)[Idle] > IMPL(v_es1)[f1])))
        |  (  EXISTS f1: field
            (  FORALL f2: field
              (  f1 ≠ f2
                →    (  IMPL(v_es1)[f1] > IMPL(v_es1)[f2]
                      &  IMPL(v_es2)[f1] > IMPL(v_es2)[f2])))))

This mapping states that two constants of type Enabled_State in the upper level are equal if and only if either (1) all the fields in the structure generated from the implementation of one of the constants are zero and the other structure is either all zeroes or the Idle field is greater than all the other fields or (2) there is a field that is greater than all the other fields in both structures in the lower level.

For the implementation of Phone_State, each server has a variable f(phone): time, for each field f of the IMPL(Enabled_State) structure that the server is responsible for. The mapping for Phone_State is shown below.

```
IMPL(Phone_State(P)) ==
     choose v_es: IMPL(Enabled_State)
        (    v_es[Idle] = TC.Idle(P)
         &   v_es[Ready_To_Dial] = GDT.Ready_To_Dial(P)
         &   ...)
```

This mapping specifies that the state of a phone P is determined by the server that has most recently timestamped a field of P. Thus, it is possible for all the servers to directly affect the state of a phone. Figure 8 illustrates an example Phone_State mapping. In this figure, each component of the state is managed by a different process and the current state of the phone is "ringing" because the corresponding timestamp component holds the maximum value.

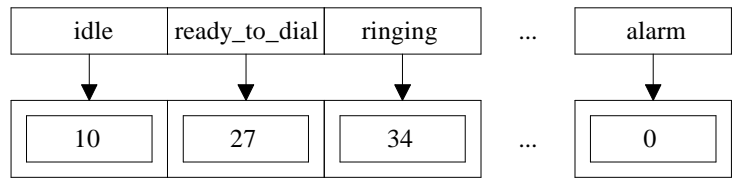

Figure 8: Implementation of the phone state

Figure 9 shows the mapping from the servers of the lower level to the values of Enabled_State. Note that the value "Calling" is not mapped to any server because Calling is only used for long distance calls. For this mapping to work, it must be guaranteed that no two servers ever give the same timestamp to the same phone. This is a problem, for example, if a phone is offhook and a "slow" server begins to serve the phone and then while this is occurring, the user of the phone hangs up, and the TC server attempts to set the phone to Idle.

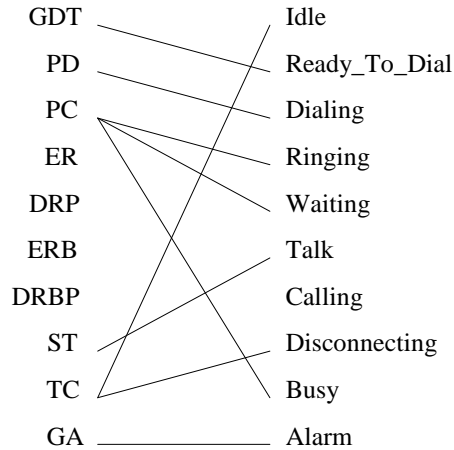

Figure 9: Mapping from servers to Enabled_State values

This problem can be avoided by keeping the Begin_Serve/Complete_Serve mechanism of the top level. Each server will only attempt to serve a phone if no other server is serving that phone. No two servers will ever be able to execute Begin_Serve at the same time for the same phone because at any given time, there is a unique function that a phone needs next. Since discrete time is assumed, it can also be guaranteed that Begin_Serve's cannot overlap on different servers. Thus, no two servers can ever give the same timestamp to the same phone.

For the implementation of serving, each server has a variable serving_set: set_of_phone. Instead of storing the value of serving for every phone, a server only needs to store those phones that it is currently serving. The mapping for serving is shown below.

      IMPL(serving(P)) == EXISTS SP: server (P ISIN SP.serving_set)

The value of K_max in the upper level is intended to limit the amount of parallelism in the system for sequential implementations. In the parallel refinement, it is undesirable to place any such limitation on the number of phones that can be served system wide. Thus, the mapping for K_max is the maximum allowable parallelism sum_K, where $\text{sum\_K} = \sum_g K\_W\_g$.

      IMPL(K_max) == sum_K

All other constants map to themselves.

In the top level, Begin_Serve is enabled when there exists a set of phones S that can be partitioned into ten disjoint subsets such that there is a subset S_g for each function g that is limited in size by K_W_g - set_size(serving_g) and must contain only the phones that begin being served. In the second level, each server has a Begin_Serve transition that is enabled when there exists a set of phones that corresponds to the disjoint subset S_g for the function g that the server performs. The exit assertion of Begin_Serve on a server SP_g specifies that the set of phones that begin being served by SP_g is equal to the disjoint subset S_g. The definition of Begin_Serve for the PD server is shown below.

```
Begin_Serve(S: nonempty_set_of_phone)
    ENTRY    [TIME:   serve_dur]
        now MOD (2 * serve_dur) = 0
      & S CONTAINED_IN W_PD
      & S SET_DIFF serving_all = S
      & set_size(S UNION serving_set) ≤ K_W_PD
      & (  set_size(S UNION serving_set) < K_W_PD
        →    W_PD CONTAINED_IN S UNION serving_all)
    EXIT
        serving_set = serving_set′ UNION S
```

A start of Begin_Serve in the upper level corresponds to a start of Begin_Serve on some server in the lower level. The mapping for an end of Begin_Serve is similarly defined.

```
IMPL(Start(Begin_Serve, now)) ==              IMPL(End(Begin_Serve, now)) ==
    EXISTS SP: server                             EXISTS SP: server
       (SP.Start(Begin_Serve, now))                  (SP.End(Begin_Serve, now))
```

The definition of Complete_Serve for servers in the lower level is similar to the definition of Complete_Serve in the upper level except that each server SP_g only checks the phones it is serving when determining which phones were being served for Dur_g and changes the state of these phones according to Exit_g. The definition of Complete_Serve for the PD server is shown below.

```
Complete_Serve
    ENTRY    [TIME:   serve_dur]
        now MOD (2 * serve_dur) = serve_dur
      & EXISTS P: phone
           (  P ISIN serving_set
           &  now - change(P ISIN serving_set) + serve_dur ≥ Dur_PD - serve_dur)
    EXIT
        FORALL P: phone
           (IF    P ISIN serving_set′
             &  now - past(change(P ISIN serving_set), now - serve_dur) ≥
                    Dur_PD - serve_dur
            THEN
                P ~ISIN serving_set
              & IF FORALL SP: server
                     (  SP ≠ GDT
                     →    GDT.Ready_To_Dial(P) > SP.ES(P))
                THEN
                    Number(P) BECOMES LISTDEF(P.Next_Digit′)
                  & Dialing(P) = now
                ELSE
                    Number(P) BECOMES
                        Number′(P) CONCAT LISTDEF(P.Next_Digit′)
                FI
            ELSE
                P ISIN serving_set′ ↔ P ISIN serving_set
            FI)
```

When the state of a phone P was previously Ready_To_Dial, the new state of P is set to Dialing, since the timestamp of Dialing(P) is set to the current time by the expression Dialing(P) = now. Note that the expression SP.ES(P) is a convenience notation to refer to the timestamp of phone P for any variables that are components of the Phone_State mapping in the server SP (e.g. TC.Idle(P)).

A start of Complete_Serve in the upper level corresponds to a start of Complete_Serve on some server in the lower level. The mapping for an end of Complete_Serve is similarly defined.

```
IMPL(Start(Complete_Serve, now)) ==         IMPL(End(Complete_Serve, now)) ==
   EXISTS SP: server                            EXISTS SP: server
      (SP.Start(Complete_Serve, now))              (SP.End(Complete_Serve, now))
```

### 9.3. Proof of Parallel Refinement of Top Level Central Control

The proof obligations that will be shown for the parallel refinement of the top level of the central control are the impl_trans_entry obligation for Begin_Serve, the impl_trans_exit obligation for Complete_Serve, and the impl_trans_fire obligation. This section discusses only the proof of the impl_trans_entry obligation for Begin_Serve, which is shown below. The other obligations and their proofs are in Appendix B. In the proof obligations, let W_g, serving_g, and serving_all refer to IMPL(W_g), IMPL(serving_g), and IMPL(serving_all), respectively.

```
FORALL t1: time
   (past
      (   EXISTS SP: server (SP.Start(Begin_Serve, now))
      →   EXISTS S: nonempty_set_of_phone
            (   now MOD (2 * serve_dur) = 0
            &   EXISTS S_GDT, S_PD, ..., S_TC: set_of_phone
                  (   S_GDT CONTAINED_IN W_GDT
                  &   S_GDT SET_DIFF serving_all = S_GDT
                  &   set_size(S_GDT UNION GDT.serving_set) ≤ K_W_GDT
                  &   ...
                  &   S = S_GDT UNION S_PD UNION ... UNION S_TC
                  &   set_size(S UNION serving_all) ≤ sum_K
                  &   (   set_size(S UNION serving_all) < sum_K
                      →   FORALL g: central function
                            (   set_size(S_g UNION SP_g.serving_set) = K_W_g
                            |   W_g CONTAINED_IN S_g UNION serving_all)))), t1))
```

Begin_Serve can only be enabled at times that are multiples of 2 * serve_dur. Complete_Serve can only be enabled at times that are serve_dur plus a multiple of 2 * serve_dur. Begin_Serve and Complete_Serve each have duration serve_dur, thus any time Begin_Serve fires on some server, every other server will either be idle or will fire Begin_Serve. At t1, Begin_Serve fires on some server, thus every other server is either idle or fires Begin_Serve. For each server SP_g on which Begin_Serve fires at t1, let S2_g be the set parameter for which Begin_Serve fired. The consequent is satisfied by the union of all sets S2_g.

The first part of the consequent is satisfied by the entry of Begin_Serve on a server on which it started. The second part is satisfied by S_g = S2_g for the servers SP_g that Begin_Serve started on at t1 and by S_g = ∅ for the other servers. By the entry of Begin_Serve, S2_g CONTAINED_IN W_g, S2_g SET_DIFF serving_all = S2_g, and set_size(S_g UNION SP_g.serving_set) ≤ K_W_g. The empty set trivially satisfies these constraints. By the definition of S, S = UNION S2_g = UNION S2_g UNION EMPTY. Each S_g UNION serving_g must be ≤ K_W_g by the entry of Begin_Serve, so set_size(S UNION serving_all) must be ≤

sum_K. Suppose set_size(S_g UNION serving_g) < K_W_g & W_g ~CONTAINED_IN S UNION serving_all for some g. Let P be a phone that needs service from g, but is not in S_g UNION serving_g. Suppose SP_g starts Begin_Serve at t1. In this case, S2_g does not satisfy the entry assertion of Begin_Serve because set_size(S2_g) < K_W_g and yet P is in W_g, but not being served by any other server. Therefore, ~SP_g.Start(Begin_Serve(S2_g), t1) holds, which is a contradiction. Suppose SP_g did not start Begin_Serve at t1. In this case, the entry assertion holds because P needs service and SP_g still has capacity left. SP_g must be idle because Begin_Serve and Complete_Serve are mutually exclusive by their entry assertions. Thus, SP_g.Start(Begin_Serve, t1) holds, which is also a contradiction.

### 9.4. Parallel Refinement of Second Level Process Call Server

This section discusses the parallel refinement of the second level process call (PC) server. The other servers of the second level central control can be refined in a similar manner. The correctness proofs for these refinements are shown in Appendix C.

The PC server is implemented by a parallel array of K_W_PC *microservers*, where each microserver is devoted to processing the calls of a single phone. Each microserver picks a phone from W_PC according to some possibly nondeterministic policy and inserts its identifier into a set of served phones through a sequence of two transitions. The union of the elements of such sets over all the PC microservers implements the serving set of the upper level PC server.

At this refinement level, it is not possible to statically allocate the individual phone timestamps of Ringing, Waiting, and Busy to different microservers or else there would be no way for phones allocated to the same microserver to be serviced at the same time, which is possible at the higher levels. Instead, microservers dynamically hold the state of the set of phones that were last serviced on that microserver. To control the size of the served set, a microserver removes a set of phones from the set such that each phone is in the set if the timestamp for that phone on some other upper level server (e.g. the ER or ERP servers) has changed more recently than the phone was added to the served set. Note that in ASTRAL, each process type is refined independently; thus, lower level processes are specified as if the other upper level processes (i.e. the processes not being refined) have not changed. Therefore, the PC microservers here refer to the timestamps of the other upper level servers as opposed to the timestamps of microservers of other function types (e.g. ER or ERP microservers).

The PC microservers process phones in pairs, where each pair is of the type phone_pair: STRUCTURE OF (Waiting: phone, Ringing: phone). The variable serving: boolean specifies if the microserver is currently serving a pair of phones. The variable serving_pair: phone_pair specifies the pair of phones the microserver is going to connect. Finally, the variable served_set: list of phone_pair specifies the set of phone pairs whose calls have been processed by the microserver, but whose state has not yet been changed by any of the other upper level servers.

The waiting timestamp for a phone P is the time that P became the waiting phone of a phone pair in the served_set of some microserver. The waiting timestamp of P is zero if no such phone pair exists.

```
IMPL(Waiting(P)) ==
    IF    EXISTS MSP: microserver, PP: phone_pair
              (   PP ISIN MSP.served_set
              &   PP[Waiting] = P
              &   PP[Ringing] ≠ P)
    THEN
          change(EXISTS MSP: microserver, PP: phone_pair
                    (   PP ISIN MSP.served_set
                    &   PP[Waiting] = P
                    &   PP[Ringing] ≠ P))
    ELSE
          0
    FI
```

The ringing timestamp for a phone P is the time that P became the ringing phone of a phone pair in the served_set of some microserver. The ringing timestamp of P is zero if no such phone pair exists.

```
IMPL(Ringing(P)) ==
    IF    EXISTS MSP: microserver, PP: phone_pair
              (   PP ISIN MSP.served_set
              &   PP[Ringing] = P
              &   PP[Waiting] ≠ P)
    THEN
          change(EXISTS MSP: microserver, PP: phone_pair
                    (   PP ISIN MSP.served_set
                    &   PP[Ringing] = P
                    &   PP[Waiting] ≠ P))
    ELSE
          0
    FI
```

The busy timestamp for a phone P is the time that P became both the waiting phone and the ringing phone of a phone pair in the served_set of some microserver. The busy timestamp of P is zero if no such phone pair exists.

```
IMPL(Busy(P)) ==
    IF    EXISTS MSP: microserver, PP: phone_pair
              (   PP ISIN MSP.served_set
              &   PP[Waiting] = P
              &   PP[Ringing] = P)
    THEN
          change(EXISTS MSP: microserver, PP: phone_pair
                    (   PP ISIN MSP.served_set
                    &   PP[Waiting] = P
                    &   PP[Ringing] = P))
    ELSE
          0
    FI
```

A phone is connected to a phone P if it is in a phone pair with P on some microserver. Connected_To is set to P otherwise.

```
IMPL(Connected_To(P)) ==
    IF    EXISTS P2: phone
              (  P2 ≠ P
              &  EXISTS MSP: microserver, PP: phone_pair
                     (  PP ISIN MSP.served_set
                     &  (  PP[Waiting] = P & PP[Ringing] = P2
                        |  PP[Waiting] = P2 & PP[Ringing] = P)))
    THEN
        choose P2: phone
              (  P2 ≠ P
              &  EXISTS MSP: microserver, PP: phone_pair
                     (  PP ISIN MSP.served_set
                     &  (  PP[Waiting] = P & PP[Ringing] = P2
                        |  PP[Waiting] = P2 & PP[Ringing] = P)))
    ELSE
        P
    FI
```

The implementation of the serving set is the set of phones that are the waiting phone in the serving pair of some microserver that is serving.

```
IMPL(serving_set) ==
    setdef P: phone
        (EXISTS MSP: microserver
            (  MSP.serving  &   MSP.serving_pair[Waiting] = P))
```

Each PC server has two transitions, which correspond to Begin_Serve and Complete_Serve in the upper level. Begin_Serve finds a pair of phones in W_PC to be connected. Complete_Serve commits the connection between the two phones identified in the preparation phase and resets the state of all phones in its list Served_Phones that have been serviced more recently by some other upper level server. The definitions of Begin_Serve and Complete_Serve are given below.

```
Begin_Serve(P: phone)
    ENTRY    [TIME:   serve_dur]
        now MOD (2 * serve_dur) = 0
        & ~serving
        & P ISIN W_PC
        & P ~ISIN serving_all
        & set_size(setdef P2: phone (P2 ISIN W_PC & P2 ~ISIN serving_all & P2 < P))
            = set_size(setdef MSP: microserver (~MSP.serving & MSP < self))
    EXIT
        serving
        & serving_pair = choose PP: phone_pair
                            (  PP[Waiting] = P
                            &  PP[Ringing] =
                                IF    GET_ID(PD.Number(P)).Offhook
                                   |  EXISTS SP: server
                                          (  SP ≠ TC
                                          →  SP.ES(GET_ID(PD.Number(P))) >
                                                  TC.Idle(GET_ID(PD.Number(P))))
                                   |  EXISTS g: central function
                                          (GET_ID(PD.Number(P)) ISIN W_g)
                                THEN
                                    P
                                ELSE    GET_ID(PD.Number(P))    FI)
```

The last conjunct of the entry assertion states that if there are n phones satisfying the condition whose IDs are less than P's ID, then there exist n microservers whose IDs are less than self and are available. This is a simple trick to state that the available microservers are allocated in order of increasing ID number to phones that need their service. In this way, conflicts are avoided and it is possible to easily prove requirements on the number of phone calls that will be served. A start of Begin_Serve in the upper level corresponds to a start of Begin_Serve on some microserver in the lower level. The mapping for an end of Begin_Serve is similarly defined.

```
IMPL(Start(Begin_Serve, now)) ==           IMPL(End(Begin_Serve, now)) ==
    EXISTS MSP: microserver                    EXISTS MSP: microserver
        (MSP.Start(Begin_Serve, now))              (MSP.End(Begin_Serve, now))
```

Complete_Serve finishes serving the phones in serving_pair.

```
Complete_Serve
    ENTRY     [TIME:   Dur_PC - serve_dur]
        serving
    EXIT
        FORALL PP: phone_pair
            (IF    PP ISIN served_set'
             &  EXISTS SP: server
                    (SP.ES(PP[Waiting]) >
                        past(change(PP ISIN served_set),
                            now - Dur_PC + serve_dur)
             &  EXISTS SP: server
                    (SP.ES(PP[Ringing]) >
                        past(change(PP ISIN served_set),
                            now - Dur_PC + serve_dur)
            THEN
                PP ~ISIN served_set
            ELSE
                PP ISIN served_set' ↔ PP ISIN served_set
            FI)
```

Notice that the duration of Complete_Serve is now Dur_PC - serve_dur, which is the time it takes to complete processing a call, whereas in the higher levels, the duration was a small duration, serve_dur, so that phones could complete being serviced at almost any time. Also note that it should be possible to place a bound on the number of phones that any microserver actually has to delete from its serving list in any execution of Complete_Serve as well as the size of the served set based on Dur_g and K_W_g of all upper level servers g.

## 10. Parallel Refinement Guidelines

During the application of the parallel refinement mechanism to different systems, several patterns emerged that are relevant to all such systems. Of utmost importance is the proper specification of the top level process. Lower level implementations must behave within the constraints imposed by the upper level. The top level process should therefore be specified so as not to impose restrictions on the overlap of activities that would disallow asynchronous concurrency in lower levels. The amount of concurrency must be bounded, however, to allow for purely sequential implementations. When concurrency is applied at lower levels, a key restriction in the ASTRAL model is that only a single process may write a given variable. Through the aggregation of individual timestamped variables, however, multiple lower level parallel processes may all effectively write to the same variable. The set of design guidelines extracted from these patterns are presented in this section.

**10.1. Asynchronous Concurrency**

Each process in a system performs some set of actions during its execution. In the implementation of a process, it may be neither feasible nor desirable for lower level processes to execute these actions in a lockstep fashion. Instead, lower level processes may need to perform actions dynamically and without synchronization with other lower level processes.

In order to allow such asynchronous concurrency in the refinement of a process, the upper level process needs to be specified appropriately. In particular, concurrent actions in the upper level that may be executed asynchronously in the lower levels should not be specified such that they begin and complete execution in the same transition. For example, in the phone system, the top level could have been specified such that there was a single transition Serve that was executed every t time units in which some set of phones was completely serviced in each execution. This would mean, however, that in the lower levels, phones could only be serviced at the rate of the slowest server and that the servers would process phones in lockstep with each other.

To allow asynchronous concurrency, concurrent actions in the upper level should be specified such that a set of actions can start and a set of actions can end at every time in the system. For example, in the top level of the central control, the service of a phone was split into the beginning of servicing and the completion of servicing in the transitions Begin_Serve and Complete_Serve, respectively. The durations of Begin_Serve and Complete_Serve were set to serve_dur, where 2 * serve_dur was chosen to be a divisor of the duration of every action. In general, it is not necessary to have a separate transition for the beginning and completion of an action. It is necessary, however, to have some record of when an action has started so that it can be completed at the appropriate time. In the central control, changes to the serving variable were used to record this information. When serving changed to true for a phone at time t, that phone began being served at t - serve_dur. Thus, when the duration of the function that was serving the phone elapsed, the effect of the function was carried out on the phone's state and serving for that phone was reset to false.

**10.2. Sequential Implementations**

In some cases, such as in the central control, there is the possibility that a process may be implemented in both a sequential and a parallel fashion. In these cases, it is necessary for the upper level specification to allow the possibility of multiple actions occurring at the same time and yet not actually requiring multiple actions to occur.

In the top level of the central control, this was achieved by the K_max constant. The K_max restriction in the entry assertion of Begin_Serve limits the number of phones that can be serviced at any given time in the system. In the sequential refinement, K_max was set to one, indicating that only one phone at a time can be serviced. In the parallel refinement, K_max was set to the sum of the capacities of the individual servers, indicating that as many phones as is possible for the servers to serve can be serviced in parallel.

When there is a nondeterministic choice of actions in the upper level, it is necessary to make the choice of actions as a transition parameter in order to allow a sequential refinement of the process. For example, in the top level of the central control, the choice of phones to begin serving was made at the start of Begin_Serve as the set parameter S. This choice could also have been made by a nondeterministic choose expression in the exit assertion of Begin_Serve. This would not have allowed for a sequential refinement of the top level, however, because in the sequential refinement of the central control, as soon as any transition begins execution, the phone and the service to be performed on the phone is immediately known. If the choice of phones and services in the top level was made in the exit assertion, it would not have been possible to determine which phone was going to be served until serve_dur after a phone actually started being served in the top level.

**10.3. Multiple Writers**

In the design of complex systems such as the phone system, there is often a need in the lower levels for multiple processes to control the implementation of a particular upper level variable. In the ASTRAL model, however, only a single process can change the value of a variable, thus it is not possible to let multiple lower level processes change the same variable directly. In the refinement of the central control, the Phone_State variable of the upper level needed to be changed by many of the servers. The solution used in that refinement, which will work in general for any refinement in which multiple writers need to be allowed, was to split the variable into a structure of timestamps, with one timestamp allocated to each process that needs to change the variable.

The Enabled_State type was simple because there were a finite number of values and each value was the responsibility of a single process. In general, however, the same technique can be used for types with arbitrary values and with an arbitrary number of writers of each value. Consider a variable v of type integer in the upper level. Suppose there are n processes P1 ... Pn that need to change the value of the implementation of v in the lower level to an arbitrary value. In order to specify this, each process Pi has a variable tv of type STRUCTURE OF (timestamp: time, value: integer). Whenever Pi changes the value of tv[value], it sets tv[timestamp] to now.

The mapping for v would then be:
```
IMPL(v) ==
        choose val: integer
          (EXISTS P: Pi
             (   P.tv[value] = val
             &   FORALL P2: Pi
                   (P ≠ P2 → P.tv[timestamp] > P2.tv[timestamp])))
```
This states that the value of v is the value of tv[value] of the lower level process that has last changed its tv. Thus, each Pi only changes its own variables and yet the implementation of v can effectively be changed by any Pi.

## 11. Related Work and Discussion

The notion of refinement is arguably one of the cornerstones of the whole discipline of computer science. Indeed, it has a long history, entangled with that of its "complement": abstraction [Dij76, Win06, Kra07]. As it is inevitable for any broad and fruitful principle, several variants have been developed under the same name, both in methodological and in technical flavors. Therefore, without any aim at comprehensiveness, in this section we sketch a "map" of the most significant of these variants, and we highlight those that are most similar to the idea of refinement that we pursue in this paper. Then, we also compare our approach with the closest ones to our own, in particular those dealing with the notion of parallelism for real-time systems.

First, let us point out that there are two main "views" usually given to the notion of refinement. The first one prescribes that a refinement B of a module A must preserve *all observable properties* of A, that is all properties that concern externally visible entities. The second notion of refinement, instead, prescribes that a refinement B of a module A must show a *subset of all observable behaviors* of A. Therefore, in general, B will have fewer possible behaviors; hence, this notion of refinement can be regarded as a lessening of the nondeterminism of a system. Having drawn this distinction, we now briefly present some refinement frameworks, ordered according to the class of program (or system) model: from sequential, to concurrent (or parallel), to real-time. More emphasis is given to the approaches that are more similar to our parallel refinement framework.

**11.1. Refinement for sequential programs.**

For descriptive formalisms such as Z and B a rich theory of sequential refinement is available, mostly based on some form of calculus, such as refinement calculus [Bac81, Bac88, Mor87, Mor94]. Some works have also argued that the refinement paradigm is too strict to deal with the modeling and development problems faced with complex and heterogenous systems. In particular, retrenchment [BP98] has been proposed as a more liberal version of the classical refinement theory that nonetheless retains some structure in the refinement proofs. Retrenchment is more liberal than "traditional" refinement in that it allows the interface to change between the abstract and the refined module, provided the two interfaces satisfy some additional constraint that is part of the retrenchment proof obligation. Although the use of retrenchment has also been advocated for systems with concurrent and real-time components [BPJS05], it is not focused on those concerns, and indeed it does not have any peculiar feature to deal explicitly with the refinement of timing behavior.

**11.2. Refinement for concurrent systems.**

Parallel refinement consists of replacing the functionality of a module by two (or more) modules that work concurrently. In the formal methods community, refinement for concurrent systems is tightly connected to decomposition techniques based on compositionality [dRLP98, dRdBH+01, Fur05]. Also, a number of works have developed a mostly semantic theory of parallel refinement, in that they have drawn general results on refinement techniques for concurrent systems without dealing explicitly with any formal notation (e.g., Abadi and Lamport [AL91]). Then, authors have specialized these general theories for particular formalisms. For instance, Cau [Cau00] deals with refinement in a concurrency framework based on dense time and temporal logic formulas, according to Abadi and Lamport's refinement techniques.

Alur and Henzinger [AH99] provide a rich formal notation called *reactive modules*, suitable to describe concurrent systems. Reactive modules can be composed both spatially and temporally. Spatial composition involves the usual operations of variable renaming and hiding, as well as parallel composition. Temporal composition involves the operations of round abstraction and triggering. Round abstraction means that several temporally consecutive rounds are combined into a single "scaled" one; in other words, it goes in the opposite direction of refinement, as it abstracts away some fine-grain details of temporal behavior. Triggering, on the other hand, splits a round into intermediate rounds so that the module sleeps through them until some predefined variables change their values. In this respect, triggering is similar to ASTRAL's sequential composition, with the difference that the reactive module formalism does not handle precise timing information. It deals only with the notion of atomicity of a transition with respect to an outside observer.

ASM is another operational formalism for which parallel refinement techniques have been developed [Boe03]. In addition, process algebras are another large class of concurrency formalisms for which rich refinement theories have been developed. In the context of the ASM formalism, let us stress again that the overwhelming majority of works in compositional and algebraic refinement do not deal explicitly with precise timing information, but only with the relative ordering of actions, events, and state changes in the execution of a system.

**11.3. Refinement for real-time systems.**

The relevance of precise timing information is what defines real-time systems. The aforementioned refinement techniques can only provide a framework into which rules for refining real-time systems can be developed. In addition, it is crucial to supplement them with suitable refinement rules that guarantee the preservation of timing information when refining a process.

Indeed, some refinement frameworks reference some real-time formalism, but they rest on general-purpose refinement mechanisms, without providing techniques that explicitly tackle the preservation of real-time properties and timed behaviors. For instance, Sampaio et al. [SMR04] address refinement of UML-RT — one of the several semi-formal real-time extensions of UML — and a formal enrichment thereof through OhCircus [CSW 03]. However, neither the "semiformal" UML-RT, nor its formal enrichment deal explicitly with timing issues, and the refinement process exclusively concerns the traditional refinement of data types, classes and their *untimed* behaviors. Thus, the user is left without mechanisms to tackle explicitly quantitative timing constraints when applying the refinement framework.

Let us therefore focus on the few approaches dealing more explicitly with quantitative time analysis. Mahony and Hayes have extended the aforementioned refinement techniques for Z to deal with timed systems [MH92]. They do so mostly in an *ad hoc* manner, focusing on the development of case studies. Real-time properties are described by explicitly introducing a variable $t$, representing absolute time, and by relating the value of functions and predicates to the values of $t$. This solution has the advantage of being simple to implement, but it also necessarily directly exposes the user to all the subtleties and problems of reasoning about time. As a consequence the refinement techniques are mostly a modification of traditional refinement techniques, where checking that the timing information is preserved by a refinement still relies largely on the ingenuity of the developer.

Scholefield et al. [SZH94] develop a variant of refinement calculus for a process algebraic formalism called TAM (Temporal Agent Model). Their notion of refinement is based on lessening of nondeterminism, rather than on the exact preservation of observable behavior. Also, TAM is an "in-the-small" language, not endowed with modular constructs needed to build large specifications.

Schneider [Sch97] introduces a theory of refinement for a timed extension of the Communicating Sequential Processes (CSP) algebra. It introduces timing information in a specification that is originally untimed. In other words, traditional refinement techniques for CSP are revised and adapted to deal with refinement steps that introduce time.

Utting and Fidge [UF96] propose a timed refinement technique that combines some of the features of timed refinement calculi with those of the TAM model. Their approach deals with discrete time only, and it allows one to refine a high-level specification into a target programming language where readings of a real-time clock are allowed to implement real-time features. This is different than the ASTRAL approach of this paper, where the same language describes the same system, at different levels of abstraction. Also, from an environment modeling perspective, the use of clocks is only an implementation-oriented way of enforcing real-time constraints, whereas a more abstract, and "physical", view of quantitative time is usually preferable during the earliest modeling phases in system development [Jac06].

Broy [Bro01] studies the notion of timed refinement with respect to the semantic model of streams. The approach is rather general, as it allows both discrete and dense time domains, and introduces a general notion of composition that permits both sequential and parallel compositions as special cases. The abstract framework is mostly concerned with general definitions and problem framing, rather than practical and readily available solutions. In this sense, it is rather orthogonal to the focus of the present paper.

As we already mentioned, the results of this paper build upon the work in [CKM95], which has already been discussed in Sections 3 and 4. Another approach that is close to this paper's is presented in [RM04], where an approach for deriving a high level architectural design of real-time distributed systems from formal specifications written in the TRIO language is suggested. The approach is based on the CORBA standard and on its extension and formalization through TRIO. Unsurprisingly, it exhibits some mutual influences with the approach presented in this paper (recall the origins of the ASTRAL language given in Section 1); its notion of refinement, however, is much less structured than the approach presented in this paper and, therefore, implies much more work in the formal analysis of lower levels of the development process.

So far, we have discussed mostly works dealing with descriptive formalisms. However, there are also a number of approaches that investigated timed refinement for operational formalisms. In particular, timed automata of various kinds have introduced and studied notions of refinement (e.g., [LSV03, AGLS01, CK05, Fre06]). Usually, the refinement of an automaton is meant to be a specialization of its behavior with respect to a given property, rather than a transformation that preserves all observable behaviors. This usually leads to simpler theories of refinement, if compared against those for descriptive formalisms, and more focused on verification of system properties, rather than full-fledged system development.

Finally, a dual-language approach can also be pursued in developing a refinement framework. Within the relatively small pool of refinement for real-time formalisms, Ostroff's dual language framework is likely one of the most comprehensive [Ost99]. It is based on the TTM/RTTL framework, where modules are described as TTM transition systems, and specifications are written using the real-time temporal logic RTTL. The framework also has tool support. The analysis of module refinement is based on bisimulation, and it aims at formulating conditions under which we can replace a module's body by another one, without affecting the observed behavior at the interface. Ostroff's methodology is rather comprehensive and encompasses various techniques in a unified framework. Its main limitations are, on the one hand the fact that it considers only discrete time and that the use of a tick transition to model time lapsing is rather unnatural and brings a description of the evolution of the state of the system which is not intuitively appealing. In particular, this implies the introduction of a "synchrony" in system temporal descriptions, which may hinder an effective modeling of purely asynchronous processes. In addition, the refinement mechanisms based on bisimulation require a one-to-one mapping between observable entities (i.e., interface variables) in the unrefined and refined modules. This lack of flexibility (mainly a result of the traditional notion of "algebraic" refinement) may hinder the use of refinement techniques in timed systems where the implementation of timing constraints is likely to require a considerable restructuring of an abstract specification.

Another dual-language refinement framework is the one by Felder et al. [FGM 98], which addresses the issue within the context of timed Petri nets and the TRIO language. A system is modeled as a timed Petri net and its properties are described as TRIO formulas. Next, mechanisms are given that refine the original net into a more detailed one that preserves the original properties. The approach is limited, however, by the expressive power of pure Petri nets, which do not allow one to deal with functional data dependencies.

**11.4. Assessment of the parallel refinement result.**

Let us discuss how our notion of parallel composition is more general than "traditional" parallel composition of processes (as in process algebras) and analyze to what extent our framework can be generalized and applied to formalisms other than ASTRAL.

Concerning generality, "traditional" refinement is based on one-to-one mappings between interface variables. This may be too restrictive in some cases. In fact, several works advocate for more liberal approaches to refinement techniques. To the extreme, a straightforward way to implement such a liberal approach is to introduce a very loose refinement framework, namely one where proving that (timing) properties are retained in the refined module is left entirely to the developer. This is what is pursued, for instance, in pure TRIO language [MS94].

It is therefore natural to pursue a more "in-between" solution, which is a refinement framework that, while flexible, is readily applicable in a significant set of cases without the need to "re-invent the wheel". In this vein, some authors have introduced *retrenchment*, which we briefly presented earlier in this section.

Our parallel refinement technique also pursues an approach to refinement that is more flexible than the traditional one, while retaining considerable structure and constraints to make its application routine. It introduces more flexibility by distinguishing between the start and stop transitions of a process, and their calls from an external client. The call interface, as in traditional refinement, cannot be changed by the refinement. However, this aspect only describes the invocation of a process, not the details of its functioning. In particular, what is most relevant from our point of view, i.e., timing information, is not restricted directly by the call interface. On the contrary, what happens between the start and stop transitions can be changed much more liberally during parallel refinement. Thus, one may introduce considerable changes to the implementation of a module, in particular by allowing the concurrent asynchronous execution of sub-modules that are not required to synchronize at the beginning and at the end of the whole module execution. In other words, whereas "traditional" parallel composition usually requires the modules executing in parallel to share their start and stop transitions, our parallel composition mechanisms are more flexible, in that they allow the user to tune much more precisely to what extent the parallel execution is asynchronous and to what extent it requires synchronization.

We believe the feature of distinguishing between the start and stop transitions of a process, and their calls from an external client is one of the main contributions of this paper and can be applied in general to formalisms other than ASTRAL; thanks to it we have been able to develop the complex case study of the phone system in a way that would be hard to deal with using the other approaches surveyed in this section.

Of course, it is possible to devise methods to implement our notion of parallel refinement using "traditional" parallel composition. For instance, and very sketchily, if module A and B execute in parallel, but module B has to start 5 seconds after module A has started, one may simply start A and B concurrently, and then delay B by 5 seconds, before letting it performing the "real" computation. However, this approach is less natural and more tricky, and the corresponding system may be more difficult to be analyzed and implemented.

As we have discussed in the previous section, the additional flexibility of our method comes at the price of additional complexity in the refinement proof obligations. This is clearly inevitable; however, the additional complexity is proportional to the additional flexibility that is employed in each case. This has been demonstrated in Section 9, where the refinement and hierarchical proof method has been applied to a complex, realistic case, such as the phone system, which breaks the barrier of so called "toy examples". The case study also allowed the derivation of methodological guidelines (collected in Section 10) that should help potential users in the autonomous application of the method to their own projects. On the other hand, we should acknowledge that there is no "silver bullet" in our proposed method as the development and analysis of a complex system still requires a complex formalization and complex proofs.

Finally, one could even argue that in some cases there is no real decrease of complexity in exploiting a hierarchical refinement-based approach with respect to a "flat" proof of correctness of the lowest level implementation against original high-level requirements. We observe, however, that even in such exceptional cases there is a high added value in the exploitation of a layered approach in terms of reusability and maintainability. For instance, imagine that a first realization of a system is done by means of a mono-processor so that the final executable code is a purely sequential program. Later, the designers decide to move to a multiprocessor, multi-process architecture to exploit advances in hardware technology. In such a case, only a fraction of the original design and verification has to be redone. Furthermore, the critical points that are affected by the changes are clearly marked, so that there is no risk, after having replaced the changed parts of the implementation and related proofs, that hidden side effects remain uncaught.

## 12. Conclusions

ASTRAL aims to provide a disciplined and well-structured way of developing real-time systems. In fact, it was originally designed [GK 91] by joining ASLAN's [AK 85] focus on developing provably correct design from formal specifications with TRIO's [GMM 90] orientation towards formalizing timing properties of the systems. Subsequently, ASTRAL's development stressed modularization and incremental development through several refinement levels. [CKM94] developed a method to structure a system as a collection of processes and to structure the overall system's correctness proof accordingly, in such a way that its complexity can be broken into smaller pieces. This type of structuring is called *intra-level* since the system and its analysis are partitioned into several interacting components at the same level of abstraction. [CGK 97] introduced bottom-up composition of ASTRAL specifications. [CKM 95] instead addressed, in a fairly preliminary way, the issue of top-down development of a system as a hierarchy of layers obtained through several refinement steps, by following an approach that is fairly well established for traditional sequential systems (see e.g., [Abr 96]), but new challenges are encountered when attempting to extend it to concurrent, hard real-time systems. By contrast with the intra-level approach, refinement and proofs through several layers of abstraction are called *inter-level*.

This paper is, in some sense, the "conceptual closure" of a complete ASTRAL-based methodology aimed at developing complex real-time systems by exploiting modularization and refinement in a disciplined and provably correct, yet flexible and general, way. In particular, it completes the inter-level refinement mechanisms and the related proof system.

A key feature of the proposed mechanisms is the exploitation of parallelism so that global actions may be described at a high level of abstraction as individual transitions that can be refined in a lower level as several concurrent and cooperating activities. Our approach allows more generality and flexibility than the few independent ones available in the literature, which are more algebraic and synchronous in nature as was discussed in Section 11. In particular, we are not aware of an approach that allows the user to formally specify high level specifications to be refined *and incrementally proved correct* in a fully asynchronous way as shown in the case study of Section 9.

### Acknowledgments

The authors would like to thank Klaus-Peter Loehr for his participation in the development of the parallel mechanisms and Giovanna Di Marzo for her valuable comments. This research was partially supported by NSF Grant No. CCR-9204249 and by the Programma di scambi internazionali of the CNR.

# Appendix

## A. Sequential Refinement of Top Level Central Control

After redefining the top level specification of the central control, it becomes possible to show (assuming discrete time) that the original central control specification in [CGK 97] without long distance is one possible second level implementation of the top level given in the previous section. Without loss of generality, it is assumed that the transitions only begin execution at times that are multiples of 2 * serve_dur. This essentially says that 2 * serve_dur is the fastest the system can recognize external changes. This can be accomplished by assuming the clause now MOD (2 * serve_dur) = 0 is conjoined to every entry assertion. The key to this refinement is mapping K_max to one. This means that only a single function can occur at any given time in the system.

*A.1. IMPL Mapping*

The IMPL mapping for the sequential refinement of the top level central control is shown below. K_max is mapped to one to force a sequential execution. The capacity of each function is mapped to one and the duration of each function is mapped to the duration of the corresponding transition. All other constants besides those below are mapped to themselves.

    IMPL(K_max) == 1
    IMPL(K_g) == 1 for all functions g
    IMPL(Dur_g) == duration of transition corresponding to function g

A phone P being served in the upper level corresponds to the time between serve_dur after the start of some service transition and just before the end of that transition. All other variables besides serving are mapped to themselves.

    IMPL(serving(P)) ==
        EXISTS tr: transition, t: time
            (    Start(tr(P), t)
             &   t + serve_dur ≤ now
             &   now < t + Duration(tr))

A start of Begin_Serve in the upper level occurs if and only if there is a start of a transition in the lower level at the same time. An end of Begin_Serve occurs if and only if there was a start of a transition serve_dur time units earlier.

    IMPL(Start(Begin_Serve, now)) ==
        EXISTS tr: transition (Start(tr, now))
    IMPL(End(Begin_Serve, now)) ==
            now ≥ serve_dur
         &  EXISTS tr: transition (Start(tr, now - serve_dur))

A start of Complete_Serve in the upper level occurs if and only if there was a start of a transition in the lower level at a time Duration(tr) - serve_dur time units earlier. An end of Complete_Serve occurs if and only if there is an end of a transition at the same time.

```
IMPL(Start(Complete_Serve, now)) ==
    EXISTS tr: transition
        (   now ≥ Duration(tr) - serve_dur
        &   Start(tr, now - Duration(tr) + serve_dur))
IMPL(End(Complete_Serve, now)) ==
    EXISTS tr: transition (End(tr, now))
```

*A.2. Proof of Sequential Refinement*

The most interesting proof obligations in the sequential refinement of the top level central control are the impl_trans_entry and impl_trans_exit obligations. In these proof obligations, let W_g, serving_g, and serving_all refer to IMPL(W_g), IMPL(serving_g), and IMPL(serving_all), respectively.

*A.2.1. Impl_trans_entry Obligation*

In the Begin_Serve case, it must be shown that the following formula holds.

```
FORALL t1: time
    (past
        (   EXISTS tr: transition (Start(tr, now))
    →       EXISTS S: nonempty_set_of_phone
            (   now MOD (2 * serve_dur) = 0
            &   EXISTS S_GDT, S_PD, ..., S_TC: set_of_phone
                (   S_GDT CONTAINED_IN W_GDT
                &   S_GDT SET_DIFF serving_all = S_GDT
                &   set_size(S_GDT UNION serving_GDT) ≤ 1
                &   ...
                &   S = S_GDT UNION S_PD UNION ... UNION S_TC
                &   set_size(S UNION serving_all) ≤ 1
                &   (   set_size(S UNION serving_all) < 1
                    →   FORALL g: central function
                        (   set_size(S_g UNION serving_g) = 1
                        |   W_g CONTAINED_IN S_g UNION serving_all)))), t1))
```

By the antecedent, there is some transition tr_g that starts at time t1. Let P be the phone that tr_g is servicing. The existential clause of the consequent is satisfied by the set consisting of only P. By previous assumption, the transitions can only start at times that are multiples of 2 * serve_dur, thus the first conjunct of the consequent holds.

The implementation of serving only holds when a transition is in the middle of execution and serve_dur has elapsed since the transition fired. By trans_mutex, there can only be one such transition. The only transition in the middle of execution is tr_g and at t1, serve_dur time has not yet elapsed. Therefore, set_size(serving_all) = 0. The second conjunct of the consequent is satisfied by the collection of sets S_h, where S_h contains only P for h = g and is empty otherwise by the entry assertion of tr_g.

In the Complete_Serve case, it must be shown that the following formula holds.

```
        FORALL t1: time
           (past
              (   EXISTS tr: transition
                     (   now ≥ Duration(tr) - serve_dur
                     &  Start(tr, now - Duration(tr) + serve_dur))
           →  (   now MOD (2 * serve_dur) = serve_dur
              &  EXISTS P: phone, tr1: transition
                     (   P ISIN serving_g
                     &  now - change(
                              EXISTS tr: transition, t: time
                                 (   Start(tr(P), t)
                                 &  t + serve_dur ≤ now
                                 &  now < t + Duration(tr))) + serve_dur ≥
                                          Duration(tr1) - serve_dur)), t1))
```

By previous assumption, transitions only start at times that are multiples of 2 * serve_dur and have durations that are multiples of 2 * serve_dur, thus t1 - Duration(tr) is a multiple of 2 * serve_dur and t1 - Duration(tr) + serve_dur MOD (2 * serve_dur) = serve_dur. Therefore, the first conjunct holds.

Let tr_g be the transition that fires at t1 - Duration(tr_g) + serve_dur. Let P be the phone that tr_g is servicing. At t1, tr_g has not yet ended and a serve_dur has elapsed since tr_g began, thus the first part of the existential clause holds.

The implementation of serving changes whenever serve_dur has elapsed since the start of a transition or at the end of a transition. Since tr_g is still executing, the last change is at the start time of tr_g + serve_dur or t1 - Duration(tr_g) + 2 * serve_dur. Thus, t1 - (t1 - Duration(tr_g) + 2 * serve_dur) + serve_dur ≥ Duration(tr_g) - serve_dur since Duration(tr_g) - serve_dur ≥ Duration(tr_g) - serve_dur. Thus, the second part of the existential clause holds.

*A.2.2. Impl_trans_exit Obligation*

In the Begin_Serve case, it must be shown that the following formula holds.

```
             FORALL t1: time
                (past
                   (   now ≥ serve_dur
                    &  EXISTS tr: transition (Start(tr, now - serve_dur))
              →    EXISTS S: nonempty_set_of_phone
                      (FORALL P: phone
                         (IF P ISIN S
                          THEN
                                EXISTS tr: transition, t: time
                                   (   Start(tr(P), t)
                                    &  t + serve_dur ≤ now
                                    &  now < t + Duration(tr))
                          ELSE
                                EXISTS tr: transition, t: time
                                   (   Start(tr(P), t)
                                    &  t + serve_dur ≤ now
                                    &  now < t + Duration(tr))
                          ↔ EXISTS tr: transition, t: time
                                   (   past(Start(tr(P), t), now - serve_dur)
                                    &  t + serve_dur ≤ now - serve_dur
                                    &  now - serve_dur < t + Duration(tr))
                          FI)), t1))
```

By the antecedent, there is some transition tr_g that starts at time t1 - serve_dur. Let P be the phone that tr_g is servicing. The existential clause of the consequent is satisfied by the set consisting of only P. Only one phone can satisfy the setdef predicate in the consequent. P satisfies the predicate for transition tr_g and time t1 - serve_dur because Start(tr_g(P), t1 - serve_dur) from the antecedent, t1 - serve_dur + serve_dur ≤ t1, and t1 < t1 - serve_dur + Duration(tr_g) since Duration(tr_g) must be a multiple of 2 * serve_dur.

In the Complete_Serve case, it must be shown that the following formula holds.

```
        FORALL t1: time
            (past
                (   EXISTS tr: transition (End(tr, now))
     →      FORALL P: phone, g: central function
                    (IF    P ISIN past(serving_g, now - serve_dur)
                        &  now - past(change(
                                    EXISTS tr: transition, t: time
                                        (   Start(tr(P), t)
                                        &   t + serve_dur ≤ now
                                        &   now < t + Duration(tr))), now - serve_dur) ≥
                                            Dur_g - serve_dur
                    THEN
                            Exit_g(P)
                        &  ~EXISTS tr: transition, t: time
                                (   Start(tr(P), t)
                                &   t + serve_dur ≤ now
                                &   now < t + Duration(tr))
                    ELSE
                            EXISTS tr: transition, t: time
                                (   Start(tr(P), t)
                                &   t + serve_dur ≤ now
                                &   now < t + Duration(tr))
                    ↔   EXISTS tr: transition, t: time
                                (   past(Start(tr(P), t), now - serve_dur)
                                &   t + serve_dur ≤ now - serve_dur
                                &   now - serve_dur < t + Duration(tr))
                    FI), t1))
```

Let tr_g be the transition that ends at t1 for a phone P. There can only be one phone and transition for which the if condition is satisfied, since only one transition can be in the middle of execution at any given time. The if condition is satisfied for phone P and function g of tr_g.

At t1 - serve_dur, the last change of the implementation of serving is at t1 - Duration(tr_g) + serve_dur, so the first part of the condition holds since past(W_g, t1 - Duration(tr_g) + serve_dur - serve_dur) holds by trans_entry. The second part of the condition holds since t1 - (t1 - Duration(tr_g) + serve_dur) ≥ Duration(tr_g) - serve_dur.

P is the only phone for which the then branch must hold. The exit of tr_g holds for P by trans_exit. Since tr_g ends at t1, the implementation of serving no longer holds at t1, thus the then branch holds. For all other phones, the else branch must hold. Both existential clauses are false because no phone other than P was being serviced at t1 - serve_dur and no phone can be serve_dur into its execution at t1 since tr_g just ended at t1. Thus, the then branch holds.

**B. Parallel Refinement of Top Level Central Control**

This section shows the proofs of the obligations impl_trans_exit for transition Complete_Serve and impl_trans_fire for transitions Begin_Serve and Complete_Serve.

*B.1. Impl_trans_exit Obligation for Complete_Serve*

```
FORALL t1: time
   (past
       (   EXISTS SP: server
               (SP.End(Complete_Serve, now))
       →   FORALL P: phone, g: central function
               (IF     P ISIN past(SP_g.serving_set, now - serve_dur)
                  &    now - past(change(EXISTS SP: server (P ISIN SP.serving_set)),
                              now - serve_dur) ≥ Dur_g - serve_dur
               THEN
                       Exit_g(P)
                  &    ~EXISTS SP: server (P ISIN SP.serving_set)
               ELSE
                       EXISTS SP: server (P ISIN SP.serving_set)
                  ↔    past(EXISTS SP: server (P ISIN SP.serving_set), now - serve_dur)
               FI), t1))
```

Suppose there is some phone P and function g such that the if condition holds, but the then branch does not hold. Complete_Serve must be enabled on SP_g at t1 - serve_dur because now MOD (2 * serve_dur) = serve_dur by the entry assertion of Complete_Serve on a server on which it ended at t1, and P ISIN SP_g.serving_set at t1 - serve_dur and t1 - serve_dur - change(P ISIN serving_set) + serve_dur ≥ Dur_PD - serve_dur by the if condition. SP_g must be idle because Begin_Serve and Complete_Serve are mutually exclusive by their entry assertions. By trans_fire, Complete_Serve starts at t1 - serve_dur on SP_g, thus its exit assertion holds at t1, so the then branch holds.

For the else branch, suppose a phone P and central function g do not satisfy the if condition, but the status of P in SP_g.serving_set changes at t1. Begin_Serve and Complete_Serve are mutually exclusive, thus Complete_Serve on SP_g changes the status. Complete_Serve on SP_g can only change the status, however, when the if condition holds for P, which is a contradiction.

*B.2. Impl_trans_fire Obligation*

In the Begin_Serve case, it must be shown that the following formula holds.

    FORALL t1: time
      ( past
        (EXISTS S: nonempty_set_of_phone
          ( now MOD (2 * serve_dur) = 0
          & EXISTS S_GDT, S_PD, ..., S_TC: set_of_phone
            ( S_GDT CONTAINED_IN W_GDT
            & S_GDT SET_DIFF serving_all = S_GDT
            & set_size(S_GDT UNION GDT.serving_set) ≤ K_W_GDT
            & ...
            & S = S_GDT UNION S_PD UNION ... UNION S_TC
            & set_size(S UNION serving_all) ≤ sum_K
            & ( set_size(S UNION serving_all) < sum_K
            → FORALL g: central function
              ( set_size(S_g UNION SP_g.serving_set) = K_W_g
              | W_g CONTAINED_IN S_g UNION serving_all)))), t1)
      & FORALL t2: time
        ( t1 - serve_dur < t2 & t2 < t1
      → ~EXISTS SP: server
        (past(SP.Start(Begin_Serve, t2), t2))
      & ~EXISTS SP: server
        (past(SP.Start(Complete_Serve, t2), t2)))
  → EXISTS SP: server
    (past(SP.Start(Begin_Serve, t1))))

Let S be a set of phones satisfying the existential clause in the antecedent. Let S_g be a nonempty set of the second part of the existential clause. There must be such a set since S is nonempty and S is the union of all such sets. The entry assertion of Begin_Serve is satisfied by the set S_g on SP_g at t1. By the antecedent, no server is executing any transition at t1, thus Begin_Serve will fire on SP_g at t1 by trans_fire.

In the Complete_Serve case, it must be shown that the following formula holds.

    FORALL t1: time
      ( past
        ( now MOD (2 * serve_dur) = serve_dur
        & EXISTS P: phone, g: central function
          ( P ISIN SP_g.serving_set
          & now - change(EXISTS SP: server (P ISIN SP.serving_set)) +
            serve_dur ≥ Dur_g - serve_dur), t1)
      & FORALL t2: time
        ( t1 - serve_dur < t2 & t2 < t1
      → ~EXISTS SP: server
        (past(SP.Start(Begin_Serve, t2), t2))
      & ~EXISTS SP: server
        (past(SP.Start(Complete_Serve, t2), t2)))
  → EXISTS SP: server
    (past(SP.Start(Complete_Serve, t1))))

Let P and g be the phones satisfying the existential clause in the antecedent. Thus, P is in the serving set of SP_g at t1. Also, P has been being served for Dur_g - 2 * serve_dur. Thus, the entry assertion of Complete_Serve on SP_g holds at t1. By the antecedent, no server is executing any transition at t1, thus Complete_Serve will fire on SP_g at t1 by trans_fire.

**C. Proof of Parallel Refinement of Process Call Server**

The proof obligations that will be shown for the parallel refinement of the process call server are the impl_trans_mutex and impl_vars_no_change obligations.

*C.1. Impl_trans_mutex Obligation*

In the Begin_Serve case, it must be shown that the following formula holds.

```
FORALL t1: time
    (   past(EXISTS MSP: microserver (MSP.Start(Begin_Serve, t1)), t1)
 →      ~past(EXISTS MSP: microserver (MSP.Start(Complete_Serve, t1)), t1)
    &   FORALL t2: time
            (   t1 < t2 & t2 < t1 + serve_dur
 →              ~past(EXISTS MSP: microserver (MSP.Start(Begin_Serve, t2)), t2))
    &   FORALL t2: time
            (   t1 < t2 & t2 < t1 + serve_dur
 →              ~past(EXISTS MSP: microserver (MSP.Start(Complete_Serve, t2)), t2)))
```

Begin_Serve can only start at times that are multiples of 2 * serve_dur, by its entry assertion. Complete_Serve is enabled when serving holds. Begin_Serve sets serving and Complete_Serve resets serving so Complete_Serve can only start immediately when a Begin_Serve ends. Therefore, Complete_Serve can only start at times that are serve_dur after a multiple of 2 * serve_dur. Since a Begin_Serve starts at t1, Complete_Serve cannot have started on any microserver in the interval (t1 - serve_dur, t1 + serve_dur). Thus, the first and the third conjuncts of the consequent hold. Since Begin_Serve only starts at times that are multiples of 2 * serve_dur and t1 is such a multiple, Begin_Serve cannot start in the interval (t1, t1 + 2 * serve_dur), thus the second conjunct holds.

In the Complete_Serve case, it must be shown that the following formula holds.

```
FORALL t1: time
    (   past(EXISTS MSP: microserver (MSP.Start(Complete_Serve, t1)), t1)
 →      ~past(EXISTS MSP: microserver (MSP.Start(Begin_Serve, t1)), t1)
    &   FORALL t2: time
            (   t1 < t2 & t2 < t1 + serve_dur
 →              ~past(EXISTS MSP: microserver (MSP.Start(Begin_Serve, t2)), t2))
    &   FORALL t2: time
            (   t1 < t2 & t2 < t1 + serve_dur
 →              ~past(EXISTS MSP: microserver (MSP.Start(Complete_Serve, t2)), t2)))
```

By previous argument, Begin_Serve can only start at times that are multiples of 2 * serve_dur and Complete_Serve always starts at an end of a Begin_Serve. Since a Complete_Serve starts at t1, Begin_Serve cannot start on any microserver in the interval (t1 - serve_dur, t1 + serve_dur). Thus, the first two conjuncts of the consequent hold. Since Begin_Serve cannot start in the interval (t1 - serve_dur, t1 + serve_dur), Complete_Serve cannot start in the interval (t1, t1 + 2 * serve_dur). Therefore, the third conjunct holds.

*C.2. Impl_vars_no_change Obligation*

For the impl_vars_no_change obligation, it must be shown that the following formula holds. Note that implementation of Vars_No_Change(t1, t1) is not expanded for brevity.

```
     FORALL t1, t3: time
        (   t1 ≤ t3
        &   FORALL t2: time
               (   t1 < t2 & t2 ≤ t3
               →   ~past(EXISTS MSP: microserver
                          (MSP.End(Begin_Serve, t2)), t2))
        &   FORALL t2: time
               (   t1 < t2 & t2 ≤ t3
               →   ~past(EXISTS MSP: microserver
                          (MSP.End(Complete_Serve, t2)), t2))
   →    FORALL t2: time
           (   t1 ≤ t2 & t2 ≤ t3
           →   IMPL(Vars_No_Change(t1, t2)))))
```

The only way for the implementations of Waiting, Ringing, Busy, Connected_To, and serving_set to change is if the served_set on some microserver changes. The served_set of a microserver only changes when an end of a Begin_Serve or Complete_Serve occurs on that microserver. By the antecedent, there is no such end in the interval (t1, t3], thus the implementations of the variables do not change value in the interval.